\documentclass[useAMS,fleqn,usenatbib]{mnras}
\usepackage{amsmath}
\usepackage{amssymb}
\usepackage{amsfonts}
\usepackage{widetext}
\usepackage{graphicx}  
\usepackage{dcolumn}   
\usepackage{bm}        
\usepackage{verbatim}   
\usepackage{multirow}
\usepackage{ulem}
\usepackage{xcolor}

\usepackage{hyperref}
\hypersetup{colorlinks=true}

\usepackage{newtxtext,newtxmath}
\usepackage[T1]{fontenc}
\usepackage{ae,aecompl}

\usepackage{lineno}

\title[Separating Intrinsic Alignment Signal and Lensing Signal]{Separating the Intrinsic Alignment Signal and the Lensing Signal using Self-Calibration in Photo-z Surveys with KiDS450 and KV450 Data}

\author[Ji Yao et al.]{
	Ji Yao$^{1,2}$\thanks{E-mail: Ji.Yao@outlook.com},
	Eske M. Pedersen$^{2}$\thanks{E-mail: eske.m.pedersen@utdallas.edu},
	Mustapha Ishak$^{2}$\thanks{E-mail: mishak@utdallas.edu},
	Pengjie Zhang$^{1,3,4}$,
	\newauthor
	Anish Agashe$^{2}$,
	Haojie Xu$^{1}$,
	Huanyuan Shan$^{5}$
	\\
	\\
	$^{1}$Department of Astronomy, School of Physics and Astronomy, Shanghai Jiao Tong University, Shanghai, 200240, People's Republic of China\\
	$^{2}$Department of Physics, The University of Texas at Dallas, Dallas, TX 75080, United States of America\\
	$^{3}$Tsung-Dao Lee Institute, Shanghai 200240, People's Republic of China\\
	$^{4}$Shanghai Key Laboratory for Particle Physics and Cosmology, People's Republic of China\\
	$^{5}$Shanghai Astronomical Observatory (SHAO), Nandan Road 80, Shanghai 200030, China\\
}

\date{Accepted XXX. Received YYY; in original form ZZZ}

\begin{document}
	
	\maketitle
	
	\begin{abstract}
		To reach the full potential for the next generation of weak lensing surveys, it is necessary to mitigate the contamination of intrinsic alignments (IA) of galaxies in the observed cosmic shear signal. The self calibration (SC) of intrinsic alignments provides an independent method to measure the IA signal from the survey data and the photometric redshift information. It operates differently from the marginalization method based on IA modeling. In this work, we present the first application of SC to the KiDS450 data and the KV450 data, to split directly the intrinsic shape -- galaxy density (Ig) correlation signal and the gravitational shear -- galaxy density (Gg) correlation signal, using the information from photometric redshift (photo-z). We achieved a clear separation of the two signals and performed several validation tests. Our measured signals are found to be in general agreement with the KiDS450 cosmic shear best-fit cosmology, for both lensing and IA measurements. For KV450, we use partial (high-z) data, and our lensing measurements are also in good agreement with KV450 cosmic shear best-fit, while our IA signal suggests a larger IA amplitude for the high-z sample. We discussed the impact of photo-z quality on IA detection and several other potential systematic biases. Finally, we discuss the potential application of the information extracted for both the lensing signal and the IA signal in future surveys.
	\end{abstract}
	
	
	
	\section{Introduction}
	As the era in cosmological studies of precision cosmology is moving forward, the role of systematic effects has become into focus. Cosmic shear is one of the primary probes to put constraints on cosmological models and to test gravity theories at cosmological scales \citep{Kaiser1992,Hu1999,Heavens2000,Bacon2001,Ishak2005,Ishak2007,Joudaki2009,Weinberg2013,Ishak2006,Linder2007,Heavens2009,Dossett2011,Dossett2012,Dossett2013}. Unfortunately, cosmic shear measurements are affected by several systematic effects \citep{Erben2001,Bacon2001,Bernstein2002,Hirata2003,Heymans2004,Ishak2004,BridleKing,Faltenbacher2009}, which require detailed studies in regards to their impacts on cosmology and also methods to remove them. Currently, some tension persists between current cosmic shear surveys themselves, i.e. KiDS (Kilo Degree Survey, \cite{Hildebrandt2016,Hildebrandt2018}), DES (Dark Energy Survey, \cite{Troxel2017}), HSC (Hyper Suprime-Cam, \cite{HSC_Hamana2019,HSC_Hikage2019}), DLS (Deep Lens Survey, \cite{Jee2016}) CFHTLenS (Canada-France-Hawaii Telescope Lensing Survey, \cite{CFHTLenSHeymans}). Moreover, CMB experiment Planck \citep{Planck2018I}, focusing on the temperature fluctuations and polarizations at the early stage ($z\sim1100$), is in tension with most of the above results from cosmic shear (except for DLS, which agrees with Planck perfectly). To understand if this tension comes from some systematic effects or new physics beyond the current standard $\Lambda$CDM cosmological model, all the systematics need to be dealt with carefully.
	
	Intrinsic alignment (IA) of galaxies constitutes one of the most important systematic effects in cosmic shear studies. Because of the non-random orientations of galaxies due to local effects, such as the tidal gravitational field or the tidal torquing with angular momentum, there are additional non-vanishing correlations from the intrinsic galactic alignments that contaminate the cosmic shear measurements. For 2-point correlations, the contamination includes the GI-type of IA (which denotes to a correlation between a tangentially lensed background galaxy and an intrinsically aligned foreground galaxy \cite{Hirata2003} and the II-type of IA (which denotes two galaxies both intrinsically aligned toward the same matter structure), see for example  \citep{Catelan2001,Hirata2003,King2005,Mandelbaum2006,Hirata2007}. 
	For more information on this topic and some more recent developments on the IA problem, we refer the reader to the following reviews or papers  \citep{TroxelIshak,Kiessling2015,Kirk2015,Joachimi2015,Kilbinger2015,Blazek2017,Troxel2017,Mandelbaum2018}.
	
	To solve this IA contamination problem and obtain a cleaner signal of cosmic shear, multiple models of IA have been proposed. Such models take into consideration the local tidal field and angular momentum, redshift evolution, and the luminosity and types of galaxies, see for example \cite{Hirata2004,BridleKing,Okumura2009,Joachimi2013,Dossett2013,Joachimi2015,Krause2016,Blazek2015,Blazek2017,Chisari2017}. These models introduce IA nuisance parameters that are included along with the cosmological parameters to be constrained by cosmic shear data. This is referred to as the marginalization method, and is widely used in current cosmic shear studies, see for example  \cite{CFHTLenSHeymans,Hildebrandt2016,Troxel2017,Hildebrandt2018,HSC_Hikage2019,HSC_Hamana2019}. An alternative approach for IA mitigation is the Self-Calibration (SC) method for the 2-point IA \cite{SC2008,Zhang2010} which was also extended to the 3-points \cite{Troxel2012,Troxel2012b,Troxel2012c}. The SC method does not put strong assumptions on the underlying IA model. It uses extra observables from the same galaxy survey to statistically separate IA-related correlation signal from the lensing signal, then uses mathematical approximations to propagate and subtract the measured IA signal in the cosmic shear measurement, see further description on this process and some recent forecasts for cosmic shear surveys in \cite{Yao2017,Yao2018}. 
	
	In a previous work \citep{Yao2017}, we quantified the improvement one can obtain by using the original SC2008 \citep{SC2008} for the IA mitigation. It can significantly reduce the GI type of IA contamination resulting in a more accurate estimation of the best-fit cosmological parameters. The very first step of the SC2008 process is the separation of the IA signal from the lensing signal in a ``galaxy-galaxy lensing''-like approach (here it means the method is carried out within the same survey and same photo-z bin, without requiring the application of spectroscopic redshift). With such a measurement, the GI signal can be obtained in a way that is independent of the IA model. The application of this $C^{Ig}$ measurement can also be used in another SC2010 \citep{Zhang2010} method to self-calibrate the II type of IA signal. This method has also been applied in theoretical predictions in \citep{Yao2018} as well as verification with simulations \citep{Meng2018}. Moreover, it can also be used in CMB lensing to clean the IA contamination \citep{Troxel2014}. Therefore, the extraction of the $C^{Ig}$ signal is very important. 
	
	In this work, we present the separation between the IA-galaxy density (Ig) signal and the gravitational lensing-galaxy density (Gg) signal by applying the SC to the KiDS450 \citep{Hildebrandt2016} and KV450 \citep{Hildebrandt2018} data. We perform a theoretical analysis of how to apply SC to a galaxy shear catalog. We numerically calculate the key ingredient for IA separation, $Q_i(\ell)$ of photo-z. We take into account several bias corrections to the lensing catalog as discussed in \cite{Hildebrandt2016,Hildebrandt2018}, and measure the two 2-points correlation functions introduced by the SC method, as extra observables, and they show a clear separation. After achieving the separation of the lensing signal and IA signal, we discuss how these different types of extra cosmological information can help future cosmological studies. A separate Letter reporting the specific detection of the gravitational shear -- intrinsic shape correlation (GI) and related work can be found in \cite{PedersenetAl2019}. 
	
	This paper is organized as follows. In Section \ref{Section method} we review the SC method and made further development for the application to real observation data. In Section \ref{Section results} we apply the SC method to KiDS450 and KV450 data and obtained the separated lensing signal and IA signal. In Section \ref{Section discussion} we discuss the future usage of the obtained signal of SC, and the potential improvements. Some extensions on the future usage of the separated signal are included in the Appendix.
	
	\section{Methods} \label{Section method}
	
	\subsection{Self Calibration - the scaling relation} \label{Section scaling}
	
	The intrinsic alignment of galaxies leads to an additional term in observed galaxy shapes, besides the cosmic shear and the (shot + shape) noise, i.e., $\gamma^{\rm obs}=\gamma^{G}+\gamma^{N}+\gamma^{I}$. When cross-correlating the galaxy shapes, the noise will only correlate with itself, however, there will exist GI and II types of IA contamination in the correlation. In terms of power spectra, it can be written as \citep{BridleKing}:
	\begin{equation}
	C^{\gamma\gamma}_{ij}(\ell)=C^{GG}_{ij}(\ell)+C^{IG}_{ij}(\ell)+C^{GI}_{ij}(\ell)+C^{II}_{ij}(\ell)+\delta_{ij}C^{GG,N}_{ii}
	\end{equation}
	where $C^{\gamma\gamma}_{ij}$ denotes the cross shape-shape power spectrum between redshift bin $i$ and redshift bin $j$. $G,I,N$ denotes to gravitational lensing, intrinsic alignment, and noise, respectively.
	
	Because of symmetry, a choice of $i< j$ can be applied. We note that the SC2008 method \citep{SC2008} doesn't apply to auto-spectra (with $i=j$). By using only cross-spectra, the impact of $C^{GI}_{ij}$ and $C^{II}_{ij}$ are minimized, as they are mainly local effects and are only significant at small physical separations. This is because the intrinsic alignment of galaxies is caused by local effects, such as tidal alignment \citep{BridleKing} or tidal torquing \citep{Blazek2017}. As a result, the II correlation will only exist over small separations, and the strength of GI correlation will depend on the ordering of the redshift bins. 

	\setcounter{equation}{8}
	\begin{figure*}
		\begin{align}
		\eta_i(z_L,z_g) &=\frac 
		{2\int_{z^P_{ i, \rm min}}^{z^P_{ i, \rm max}}dz^P_{G}\int_{z^P_{ i, \rm min}}^{z^P_{ i, \rm max}}dz^P_g\int_{0}^{\infty}dz_G W_L(z_L,z_G)p(z_G|z^P_G)p(z_g|z^P_g)S(z^P_G,z^P_g)n^P_i(z^P_G)n^P_i(z^P_g)}
		{\int_{z^P_{ i, \rm min}}^{z^P_{ i, \rm max}}dz^P_{G}\int_{z^P_{ i, \rm min}}^{z^P_{ i, \rm max}}dz^P_g\int_{0}^{\infty}dz_G W_L(z_L,z_G)p(z_G|z^P_G)p(z_g|z^P_g)n^P_i(z^P_G)n^P_i(z^P_g)}  \label{eta}\\
		2 &= \frac{\int_{z^P_{ i, \rm min}}^{z^P_{ i, \rm max}}dz^P_{G}\int_{z^P_{ i, \rm min}}^{z^P_{ i, \rm max}}dz^P_g 
			n^P_i(z^P_G)n^P_i(z^P_g)}{\int_{z^P_{ i, \rm min}}^{z^P_{ i, \rm max}}dz^P_{G}\int_{z^P_{ i, \rm min}}^{z^P_{ i, \rm max}}dz^P_g 
			n^P_i(z^P_G)n^P_i(z^P_g)S(z^P_G,z^P_g)}, \label{eq:front2}
		\end{align}
		\hrulefill
	\end{figure*}
	\setcounter{equation}{1}
	
	In order to extract the IG correlation using the observable $C_{ii}^{Ig}$, a scaling relation is derived \citep{SC2008} under small-bin approximation,
	\begin{equation} \label{scaling-1}
	C^{IG}_{ij}(\ell) \simeq \frac{W_{ij}\Delta_i}{b_i(\ell)}C^{Ig}_{ii}(\ell),
	\end{equation}
	where $W_{ij}$ is the weighted lensing kernel, defined as: 
	\begin{subequations}
		\begin{equation}
		W_{ij} \equiv \int_{0}^{\infty}dz_L\int_{0}^{\infty}dz_S
		[W_L(z_L,z_S)n_i(z_L)n_j(z_S)]\label{Wij},
		\end{equation}
		$\Delta_i$ is an effective width of $i^{\text{th}}$ redshift bin, defined as:  
		\begin{equation}
		\Delta_i^{-1}\equiv \int_{0}^{\infty}n_i^2(z)\frac{dz}{d\chi}dz \label{Delta_i}
		\end{equation}
		and $b_i(\ell)$ is the galaxy bias averaged over the redshift bin. It can be inferred from the following relation:
		\begin{equation}
		C_{ii}^{gg}(\ell)\approx b_i^2(\ell) C_{ii}^{mm}(\ell) \label{b_i}
		\end{equation}
	\end{subequations}
	
	In Eqs.\, \eqref{Wij}, \eqref{Delta_i} and \eqref{b_i}, $W_L$ is the lensing kernel, which we will discuss later in Eq.\,\eqref{lensing kernal}; $z_L$ and $z_S$ are the redshift of the lens and source, respectively; $n_i(z)$ is the true redshift distribution of the $i^{\text{th}}$ tomographic bin; $\chi$ is the comoving distance; $C^{mm}_{ii}$ is the fiducial matter power spectrum; and $C^{gg}_{ii}$ is the observed galaxy-galaxy clustering power spectrum. Therefore, the task of measuring $C^{IG}$ is converted to that of  measuring $C^{Ig}$, the IA-galaxy correlation. In CMB lensing, a scaling relation similar to Eq.\,\eqref{scaling-1} can be obtained, see in \cite{Troxel2014}.
	
	\subsection{Self Calibration - the separation of $C^{Ig}$} \label{Section Ig}
	
	Here we review the separation of the $C^{Ig}$ signal from the lensing signal $C^{Gg}$ in the shape-galaxy correlation $C^{\gamma g}$. This is not only important in terms of the IA information subtracted using SC, as explained in the previous subsection, but is also potentially useful in future galaxy-galaxy lensing analysis with the separated $C^{Gg}$ signal. In this subsection, we reiterate the separation method, which was first derived in \cite{SC2008}.
	
	The separation of $C^{Ig}$ from $C^{Gg}$ is done by the virtue of their distinct dependencies on the relative line-of-sight distances of galaxies in a pair. In a pair of galaxies, let us denote the photo-z of the galaxy used for shape measurement as $z_{\gamma}^P$ and that of the galaxy used for number density measurement as $z_g^P$. The $I-g$ correlation does not depend on the ordering of galaxies in a pair along the line of sight as long as the physical distance between them is fixed. Therefore, $C^{Ig}$ remains the same for galaxy pairs with $z_{\gamma}^P > z_g^P$ and $z_{\gamma}^P < z_g^P$, given that the separation $|z_{\gamma}^P - z_g^P|$ does not change. However, on account of the geometry dependence of lensing, $C^{Gg}$ is statistically smaller for a pair with $z_{\gamma}^P < z_g^P$ than that for a pair with $z_{\gamma}^P > z_g^P$.
	We can, then, construct the following two observables:
	
	\begin{subequations}
		\begin{align} 
		C^{\gamma g}_{ii}&=C^{Ig}_{ii}+C^{Gg}_{ii}, \label{gamma-g} \\
		C^{\gamma g}_{ii}|_S&=C^{Ig}_{ii}+C^{Gg}_{ii}|_S \label{gamma-g|S},
		\end{align}
	\end{subequations}
	
	where "$|_S$" denotes only correlating the pairs with $z^P_\gamma<z^P_g$. Due to the fact that, unlike the lensing signal, the IA signal does not depend on the ordering of the source-lens pair, we can write $C^{Ig}_{ii}|_S=C^{Ig}_{ii}$, however, $C^{Gg}_{ii}|_S < C^{Gg}_{ii}$. 
	
	This drop in the lensing signal is captured by the quantity $Q_i(\ell)$, defined as:
	\begin{equation} \label{Q}
	Q_i(\ell)\equiv \frac{\textit{C}^{Gg}_{ii}|_S(\ell)}{\textit{C}^{Gg}_{ii}(\ell)}
	\end{equation}
	which can be calculated using only the redshift distribution, including the observed photo-z distribution and the estimated true redshift distribution, of the survey. The value of $Q_i$ can roughly represent the photo-z quality in the $i^{\text{th}}$ redshift bin. Based on the definition, the range will be $0<Q_i<1$, in which $Q_i\approx1$ stands for very bad photo-z quality, and $Q_i\approx0$ means spectroscopic-level photo-z quality.
	
	By combining Eq.\,\eqref{gamma-g}, \eqref{gamma-g|S} and \eqref{Q}, the separation of $C^{Ig}$ can be achieved \citep{SC2008}:
	\begin{equation} \label{scaling-2}
	C^{Ig}_{ii}(\ell)=\frac{C^{\gamma g}_{ii}|_S(\ell)-Q_i(\ell)C^{\gamma g}_{ii}(\ell)}{1-Q_i(\ell)}.
	\end{equation}
	
	In this work, we use KiDS450 and KV450 data to detect the differences in Eq.\,\eqref{gamma-g} and \eqref{gamma-g|S}, and show that SC can separate these two observables due to the physical differences between lensing and IA. Then the direct detection of Eq.\,\eqref{scaling-2} in configuration space is also achievable. We discuss the robustness of the signal separation, and the usage of the information from the separated lensing signal and the IA signal. The generalized SC signal to cosmic shear or CMB lensing is left for future studies.
	
	\subsection{$Q_i$ measurements} \label{Section Q}
	
	According to Eq.\,\eqref{scaling-2}, the estimation of $Q_i$ plays an important role in the IA signal separation. Theoretically, $Q_i$ is calculated using Eq.\,\eqref{Q}, in which
	\begin{subequations}
		\begin{align}
		C^{Gg}_{ii}(\ell)&=\int_0^\infty\frac{W_i(\chi)n_i(\chi)}{\chi^2} b_g P_\delta\left(k=\frac{\ell}{\chi};\chi\right)d\chi, \label{Gg} \\ 
		C^{Gg}_{ii}|_S(\ell)&=\int_0^\infty\frac{W_i(\chi)n_i(\chi)}{\chi^2} b_g P_\delta\left(k=\frac{\ell}{\chi};\chi\right) \eta_i(z) d\chi . \label{GgS}
		\end{align}
	\end{subequations}
	
	Here $W_i$ is the lensing efficiency function, defined as
	\begin{equation} \label{W_i}
	W_i(\chi_L)=\frac{3}{2}\Omega_m\frac{H_0^2}{c^2}(1+z_L) \int_{\chi_L}^\infty n_i(\chi_S)\frac{(\chi_S-\chi_L)\chi_L}{\chi_S}d\chi_S;
	\end{equation}
	where, $n_i(\chi)$ is the true redshift galaxy distribution in the $i^{\text{th}}$ redshift bin; $\chi$ is the comoving distance; $b_g$ is the galaxy bias; $P_\delta$ is the matter power spectrum.  Based on the definition, the "$|_S$" sign in Eq.\,\eqref{GgS} denotes only correlating the $\gamma-g$ pairs with $z^P_\gamma<z^P_g$. To fulfill this definition, a quantity $\eta_i(z)=\eta_i(z_L=z_g=z)$ for the $i^{\text{th}}$ tomographic bin is defined in \cite{SC2008} as Eq.\,\eqref{eta} shown on the top of this page, with the normalization factor 2 (in the numerator) defined as Eq.\,\eqref{eq:front2}.

	Here $z_L$, $z_g$ and $z_G$ denote to the redshifts of the lens, the galaxy count, and the shear source; $z^P$ is the photo-z and $z$ is the true redshift; $z^P_{i,\rm min}$ and $z^P_{i,\rm max}$ are the minimum and maximum photo-z in the $i^{\text{th}}$ tomographic bin; $W_L$ is the lensing kernel:
	\setcounter{equation}{10}
	\begin{equation}
	W_L(z_L,z_S)=\begin{cases}
	\frac{3}{2}\Omega_m\frac{H_0^2}{c^2}(1+z_L)\chi_L(1-\frac{\chi_L}{\chi_S}) &\text{for $z_L<z_S$}\\
	0 &\text{otherwise}
	\end{cases};\label{lensing kernal}
	\end{equation}
	$p(z|z^P)$ is the probability distribution function (PDF), which is assumed to be Gaussian, as in our previous works \citep{SC2008,Yao2017}:
	\begin{equation}
	p(z|z^P)=\frac{1}{\sqrt{2\pi}\sigma_z(1+z)}{\rm exp}\left\{-\frac{(z-z^P-\Delta_z^i)^2}{2[\sigma_z(1+z)]^2}\right\}; \label{eq: Gaussian PDF}
	\end{equation}
	$S(z^P_G,z^P_g)$ is the selection function for the ``$|_S$'' symbol:
	\begin{equation} \label{selection}
	S(z^P_G,z^P_g)=\begin{cases}
	1 &\text{for $z^P_G<z^P_g$}\\
	0 &\text{otherwise}
	\end{cases},
	\end{equation}
	which is normally referred to as the Heaviside step function. Here $n^P_i(z^P)$ gives the photo-z distribution function in the $i^{\text{th}}$ tomographic bin. The front-factor ``2'' in the numerator of Eq.\,\eqref{eta} is the normalization factor, which is given by Eq.\,\eqref{eq:front2}, according to \cite{SC2008}.
	
	From the expressions above, one can see that to calculate $Q_i(\ell)$, the values we need are the photo-z distribution $n^P_i(z^P)$, the true redshift distribution $n_i(z)$, the lensing theory (for example, in Eq.\,\eqref{W_i}), and a fiducial cosmology (for example, $P_\delta$ in Eq.\,\eqref{Gg}). In \cite{Yao2017}, we theoretically calculated the $Q_i(\ell)$ using the steps above for LSST. We showed that the value of $Q_i(\ell)$ is almost scale-independent, and we discussed that its measurement uncertainty is negligible for a survey with photo-z quality like LSST.
	
\begin{figure}
		\includegraphics[width=1.0\columnwidth]{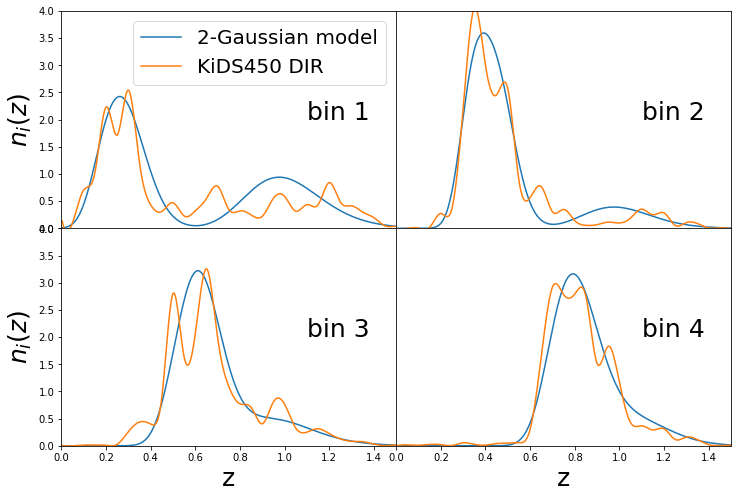}
		\caption{Here we present the redshift distribution $n_i(z)$ of our 2-Gaussian model, which is used to quantify the bias from photo-z outlier in $Q_i(\ell)$ measurement. The orange curves are the DIR $n_i(z)$ of KiDS450 \citep{Hildebrandt2016}, while the blue curves are our 2-Gaussian model with suitable scatter and outlier rate. The outlier rate ($f_{\rm out}$ in Eq.\,\eqref{2G model}) for our 2-Gaussian model is [0.4, 0.15, 0.2, 0.15]. The impact of this photo-z outlier will be propagated to the $Q_i$ measurement later in Fig.\,\ref{fig: Q_i} and the $w^{Gg}$ v.s. $w^{Ig}$ separation in Fig.\,\ref{fig: Gg and Ig} as the major systematical bias.}
		\label{fig: 2Gaussian}
	\end{figure}

	In this work, we will adopt the default Gaussian z-PDF model as in Eq.\,\eqref{eq: Gaussian PDF} to describe the characteristic behavior of the ``lensing-drop'' $Q_i(\ell)$. Meanwhile, we will also apply a bi-Gaussian model:
	\begin{equation}
	p_{\rm 2G}(z|z^P)=(1-f_{\rm out})p_{\rm main}(z|z^P)+f_{\rm out}p_{\rm outlier}(z|z^P), \label{2G model}
	\end{equation}
	to quantify the bias from non-Gaussian PDF. Each of $p_{\rm main}(z|z^P)$ and $p_{\rm outlier}(z|z^P)$ is a single-Gaussian PDF in the form of Eq.\,\eqref{eq: Gaussian PDF} but with different parameters. The quantity $f_{\rm out}$ gives the outlier rate. More details of the above bi-Gaussian model will be shown in Fig.\,\ref{fig: 2Gaussian}. However, we note this model is only a supplementary method to quantify the bias in $Q_i$, while in the main analysis we stick to the single-Gaussian model. This is for self-consistency reason: if we include the z-outlier in the PDF for $Q_i$ calculation Eq.\,\eqref{Q}, but not in the best-fit redshift being used in the selection Eq.\,\eqref{selection}, there will be some extra bias that we can not quantify.
	
	\subsection{Survey data} \label{Section data}
	
	In this work we apply the lensing-IA separation part of the self-calibration (SC, \cite{SC2008}) to Kilo Degree Survey (KiDS) data, more specifically, KiDS450 \citep{Hildebrandt2016} and KV450 \citep{Hildebrandt2018} data release's shear catalogs. The information about these two surveys can also be found here (\url{http://kids.strw.leidenuniv.nl/sciencedata.php}). KiDS is one of the main stage III surveys obtaining valuable cosmological information with weak lensing studies. 
	
	We use the shear catalog from KiDS \citep{Kuijken2015,Hildebrandt2016,FenechConti2016}. The KiDS data are processed by THELI \citep{Erben2013} and Astro-WISE \citep{Begeman2013,deJong2015}. Shears are measured using lensfit \citep{Miller2013}, and photometric redshifts are obtained from PSF-matched photometry and calibrated using external overlapping spectroscopic surveys (see \cite{Hildebrandt2016}).
	
	KiDS450 \citep{Hildebrandt2016,FenechConti2016} is a four-band imaging survey, covering $449.7{\rm \,deg}^2$ (with effective survey area $360.3{\rm \,deg}^2$) on the sky. The four bands are SDSS-like $u$-, $g$-, $r$- and $i$-bands, with corresponding limiting magnitude [24.3, 25.1, 24.9, 23.8]. The photo-z range for the galaxies in the shear catalog is ${0.1<z^P<0.9}$. Its effective galaxy number density is $\sim 8.53{\rm \,arcmin}^2$ within this photo-z range, with overall $\sim 15$ M galaxies with shear measurements. The rms shape error is $\sim 0.290$ of all the sample galaxies in the shear catalog. 
	
	KV450 \citep{Wright2018,Kannawadi2019,Hildebrandt2018} is the combination of the two sister surveys, KiDS and VIKING. It uses overall 9 photometric bands, 4 optical bands -- $u g r i$ from KiDS and 5 near-infrared bands -- $ZYJHK_s$ from VIKING. Its effective survey area reduces to $341.3{\rm \,deg}^2$ (from $360.3{\rm \,deg}^2$ of KiDS450), leading to a reduced effective galaxy number density $6.93{\rm \,arcmin}^2$ and the total number of galaxies $\sim 12$ M. The main improvement of the KV450 shear catalog is its photo-z quality \citep{Wright2018}: the photo-z scatter $\sigma_z$ improved from 0.082 (KiDS450) to 0.061 (KV450), which is more concentrated to the true redshift, while the photo-z outlier rate $\eta_3$ is reduced from 0.163 (KiDS450) to 0.118 (KV450), meaning the fraction of galaxies with catastrophic photo-z error outside $3\sigma$ range is reduced. KV450's redshift range is ${0.1<z^P<1.2}$. The rms shape error is also improved a little for KV450, with the value $\sigma_e\sim 0.288$.
	
	In this work, we have processed KiDS450 data with two pipelines (see related discussion in Appendix \ref{Apdx:Twopipelines}) to test the robustness of the method, considering their different processes for photo-z error, shape calibration, and re-sampling. As the results converge, we use only the first pipeline to process KV450 data.
	
	\subsection{Two-point statistics} \label{Section 2pt}
	
	In this section, we present the analysis of the correlation functions we need for the separation of the lensing signal and the IA signal using SC. First, we transform the target correlations $C^{Gg}_{ii}(\ell)$ and $C^{Ig}_{ii}(\ell)$ from $\ell$-space to $w^{Gg}(\theta)$ and $w^{Ig}(\theta)$ respectively, in real space. This is done using a Hankel transformation \citep{Joudaki2018}: 
	\begin{equation} \label{Hankel}
	w^{\{Gg,Ig\}}(\theta)=\frac{1}{2\pi}\int d\ell~\ell C^{\{Gg,Ig\}}J_2(\ell\theta) 
	\end{equation}
	where, the superscripts $Gg$ and $Ig$ mean shear-galaxy and IA-galaxy correlation functions respectively; and $J_2(x)$ is the Bessel function of the first kind and order 2.
	
\begin{figure}
		\includegraphics[width=1.0\columnwidth]{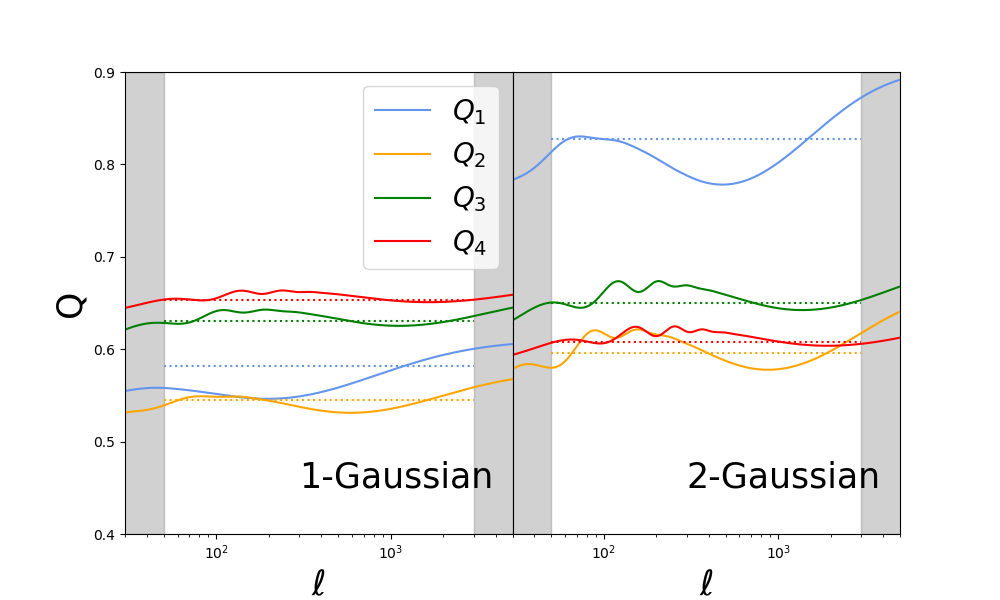}
		\caption{Here we show the $Q_i(\ell)$ calculated for the 4 redshift bins of KiDS450, with the default 1-Gaussian model and the 2-Gaussian model for comparison. The gray shaded regions are the cut in $\ell$, so that we only account for $50<\ell<3000$ in this work. The curves has weak angular dependency on $\ell$, and is nearly constant as it goes to higher redshift. The dotted lines are the mean $Q_i$ values in the 4 redshift bins, with $\bar{Q_i} \approx$ [0.58, 0.54, 0.63, 0.65] for the default 1-Gaussian model and $\bar{Q_i} \approx$ [0.83, 0.60,  0.65, 0.61] for the 2-Gaussian model. The statistical error in $Q_i(\ell)$ from the sample variance is $<\sim10^{-3}$ and is therefore negligible.}
		\label{fig: Q_i}
	\end{figure}

	As we have showed in Figs.\,\ref{fig: Q_i} and \ref{fig:Q_i_v2} here and in previous work \cite{Yao2018}, the value of $Q_i(\ell)$ remains almost constant and can be approximated to its mean value $\bar{Q_i}$. Then, using Eq.\,\eqref{Hankel} in  Eq.\,\eqref{scaling-2} gives us:
	\begin{equation} \label{Ig correlation}
	w^{Ig}(\theta)=\frac{w^{\gamma g}|_S(\theta) - \bar{Q}_i w^{\gamma g}(\theta)}{1-\bar{Q}_i},
	\end{equation}
	where, $w^{\gamma g}=w^{Ig}+w^{Gg}$ is the observed shape-galaxy correlation, in which $\gamma$ represents the observed galaxy shape $\gamma^{\rm obs}=\gamma^G+\gamma^I+\gamma^N$ as discussed earlier. 
	Nevertheless, according to Eqs.\,\eqref{gamma-g} and \eqref{gamma-g|S}, we can not only separate the IA signal, but also the lensing signal:
	\begin{equation} \label{Gg correlation}
	w^{Gg}(\theta)=\frac{w^{\gamma g}(\theta) - w^{\gamma g}|_S(\theta)}{1-\bar{Q}_i},
	\end{equation}
	which is the lensing-galaxy signal within the same $i^{\text{th}}$ redshift bin. This is different from the conventional galaxy-galaxy lensing as: (1) it correlates the lens and source within the same redshift bin in the same survey; (2) It doesn’t require highly precise redshifts, such as from spectroscopy. We will discuss this more later.
	
	In Eq.\,\eqref{Ig correlation} the observables are $w^{\gamma g}$ and $w^{\gamma g}|_S$. Therefore we need to calculate these correlation functions from the KiDS450 shear catalog. We use the following estimator \citep{Mandelbaum2006} for this purpose:
	\begin{equation} \label{gamma-g correlation}
	w^{\gamma g}=\frac{\sum_{\rm ED}w_j\gamma^+_j}{\sum_{\rm ED}(1+m_j)w_j}-\frac{\sum_{\rm ER}w_j\gamma^+_j}{\sum_{\rm ER}(1+m_j)w_j},
	\end{equation}
	where $\sum_{\rm ED}$ means summing over all the tangential ellipticity (E) - galaxy count in the data (D) pairs, $\sum_{\rm ER}$ means summing over all the tangential ellipticity (E) - galaxy count in the random catalog (R) pairs. After normalization with the number of galaxies, $\sum_{\rm ED}(1+m_j)w_j$ and $\sum_{\rm ER}(1+m_j)w_j$ are quite similar in the large scale structure we are interested in this work, as the boost factor (the ratio of these two) is 1 \citep{Mandelbaum2005boostfactor,Singh2017}.
	
	The random catalog is generated using the KiDS450 footprint mask \citep{Hildebrandt2016}, aiming to subtract the selection bias in the correlation function from the shape of the footprint of the survey. The size of the random catalog used in pipeline one is 9 times the size of the whole KiDS450 catalog.
	The large random catalog is aimed to reduce the measurement uncertainty caused by the 2nd term in the RHS of Eq.\,\eqref{gamma-g correlation}.
	
	We note in the random catalog we only included the survey geometry, however ignored the differences in observational depth in different areas on the sky. Since this approach can not capture the ``fake overdensity'' due to the different maximum depth on the sky, it will lead to some bias in the density field. This effect will also affect each redshift bin differently. We choose to ignore this effect because: (1) this ``fake overdensity'' is not related to the large scale structure, thus when cross-correlating with the galaxy shapes (which are due to the large scale structure for both lensing and IA), this effect is expected to vanish. (2) There could still exist some residual ``fake overdensity'' bias due to the selection effects, which should be, however, included in the error bars of our Jackknife resampling method. Therefore the systematical effect is transformed into the statistical effect. More detailed studies on the random catalog effects will be discussed in future work.
	
	Moreover, $w_j$ denotes the $lens$fit \citep{Miller2013} weight for the galaxy shape measurement; $\gamma^+_j$ is the tangential ellipticity for the $j^{\text{th}}$ galaxy; $(1+m_j)$ is a term used in the calibration correction for the shear measurement \citep{Hildebrandt2016}:
	\begin{equation} \label{shape correction}
	\gamma^{\rm obs}_j=(1+m_j)\gamma^{\rm true}_j+c_j,
	\end{equation}
	where the true shape of the $j^{\text{th}}$ galaxy $\gamma^{\rm true}_j$ is affected by a multiplicative bias $(1+m_j)$ and an additive bias $c_j$, and appears as the observed shape $\gamma^{\rm obs}_j$. 
	
	In cosmic shear studies, the shear-shear correlation requires subtraction of the additive bias, which can be acquired by averaging the observed ellipticity $c_j=<\gamma_j^{\rm obs}>$ \citep{Hildebrandt2016}. This is not necessary for this work as in Eq.\,\eqref{gamma-g correlation} the additive bias is already corrected with the random catalog.
	
	\subsection{(Extrapolated) Theoretical Predictions} \label{Section theory}
	
	In this section, we briefly describe how the theoretical values of $w^{Gg}$ and $w^{Ig}$ are calculated. The comparison between these theoretical values and the measurements from data will be presented in sections \ref{Section results KiDS} and \ref{Section results KV}. 
	
	The two target correlation functions are calculated from the Hankel transformation described in Eq.\,\eqref{Hankel}, in which the lensing-galaxy correlation $C^{Gg}$ is given in Eq.\,\eqref{Gg}, and the IA-galaxy correlation $C^{Ig}$ is given by
	\begin{equation}
	C^{Ig}_{ii}(\ell)=\int_0^\infty\frac{n_i(\chi)n_i(\chi)}{\chi^2}b_gP_{\delta,\gamma^I}\left(k=\frac{\ell}{\chi};\chi\right)d\chi. \label{C^Ig}
	\end{equation}
	The related 3-D matter-IA power spectrum $P_{\delta,\gamma^I}$ depends on the IA model being used. In this work, we use the tidal alignment model \citep{Hirata2004,BridleKing} for the purpose of theoretical modeling. It is the default IA model for KiDS450 \citep{Hildebrandt2016} and is commonly used in the other stage III surveys \citep{Hildebrandt2018,Troxel2017,HSC_Hikage2019,HSC_Hamana2019,Chang2019}. The form is as follows:
	\begin{equation} \label{IA 3D}
	P_{\delta,\gamma^I}=-A_{\rm IA}(L,z)\frac{C_1\rho_{m,0}}{D(z)}P_\delta(k;\chi),
	\end{equation}
	where $\rho_{m,0}=\rho_{crit}\Omega_{m,0}$ is the mean matter density of the universe at $z=0$. $C_1=5\times 10^{-14}(h^2M_{\rm sun}/Mpc^{-3})$ was used in \cite{BridleKing}. We use $C_1\rho_{crit}\approx 0.0134$ as in \cite{Krause2016}. $D(z)$ is the normalized growth factor. $A_{\rm IA}(L,z)$ is IA amplitude parameter, which is expected to be luminosity($L$)- and redshift($z$)-dependent. Here we follow the KiDS450 results \citep{Hildebrandt2016} and set it as a constant for the current stage of study.
	
	In the first pipeline, the theoretical values are calculated with CCL (Core Cosmology Library, \url{https://github.com/LSSTDESC/CCL}, \cite{Chisari2019CCL}), and cross-checked with results using CAMB (Code for Anisotropies in the Microwave Background, \url{https://camb.info/}, \cite{Lewis2000CAMB}), in the default pipeline. The cosmological parameters being used for the calculation are the best-fit values of KiDS450, shown in Table.\,\ref{table: fiducial cosmology}, with the redshift distribution of DIR redshift (\url{http://kids.strw.leidenuniv.nl/sciencedata.php}) in \cite{Hildebrandt2016}. 
	
	\begin{table}
		\caption{A list of our fiducial cosmological models, using the best-fit cosmology from different observations \citep{Hildebrandt2016,Hildebrandt2018,Planck2018I}. They are used to produce the theoretical predictions in our results.}\label{table: fiducial cosmology}
		\hspace{-0.3cm}
		\begin{tabular}{ c c c c c c c }
			\hline
			Survey & $h_0$ & $\Omega_b h^2$ & $\Omega_c h^2$ & $n_s$ & $\sigma_8$ & $w$  \\
			\hline
			KiDS450 & 0.75  &  0.0223  &  0.119  &  1.02  & 0.826 &  -1.0 \\
			KV450 & 0.745  &  0.022  &  0.118  &  1.021  & 0.836 &  -1.0 \\
			Planck2018 & 0.6732  &  0.022383  &  0.12011  &  0.96605  & 0.812 &  -1.0 \\
			\hline
		\end{tabular}
	\end{table}

	\begin{figure*}
		\includegraphics[width=2.0\columnwidth]{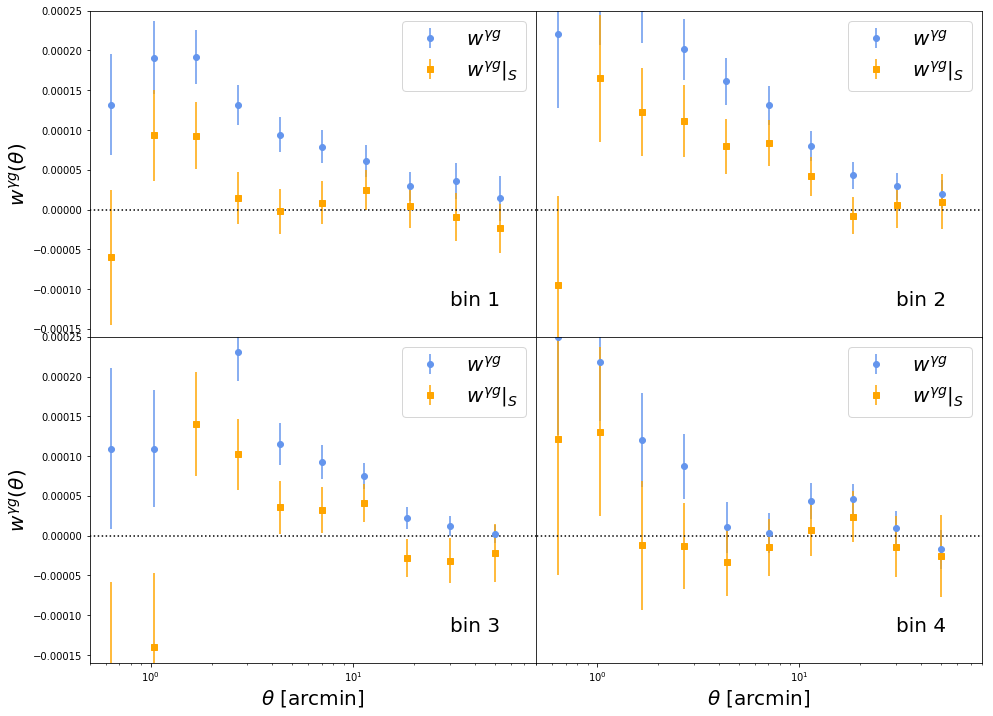}
		\caption{We show the measurements of the observables of SC, in the four redshift bins. The shape-galaxy correlations $w^{\gamma g}$ are shown in blue dots, and the shape-galaxy correlations with the selection $w^{\gamma g}|_S$ are shown in yellow squares. The redshift bin numbers from 1 to 4 are in the labels. For most of the angular bins there is a clear separation between the two correlation functions, representing the drop of the lensing signal related with Eq.\,\eqref{Q}, due to the selection of Eq.\,\eqref{selection}. The significance of the lensing-drop from $w^{\gamma g}$ to $w^{\gamma g}|_S$ for each redshift bin is [$8.7\sigma$, $8.3\sigma$, $9.0\sigma$, $4.5\sigma$].}
		\label{fig: observables}
	\end{figure*}
	
	\section{Results} \label{Section results}
	
	\subsection{$Q_i$ measurements} \label{Section results Q}
	
	We present the measurement of $Q_i$ for KiDS450 in Fig.\,\ref{fig: Q_i} and Fig.\,\ref{fig:Q_i_v2}.
	The fiducial cosmology for Eq.\,\eqref{Q} is assumed to be either KiDS450 cosmology (here) or KV450 cosmology (later in this paper) depending on which survey data we are using. The detailed values are shown in Table\,\ref{table: fiducial cosmology}.
	
	For proper usage of SC for KiDS450 data, one needs to use the photo-z distribution $n_i^P(z^P)$, in which the photo-z $z^P$ is given in the KiDS450 shear catalog as $z_B$, the peak value of the PDF (probability distribution function) from the BPZ (Bayesian Photometric redshift) code \citep{Benitez2000,Hildebrandt2016}. Not only so, we also need the full PDF $p(z|z^P)$ according to Eq.\,\eqref{eta}, as well as the true-z distribution $n_i(z)$ or $n_i(\chi(z))$ according to Eq.\,\eqref{Gg} and \eqref{GgS}. The full PDF $p(z|z^P)$ should also be given from the photo-z estimation and the true-z distribution $n_i(z)$ should be given from stacking the PDFs of all the galaxies.
	
	However, we point out that for this work, this approach is not applicable. According to \cite{Hildebrandt2016}, the stacked redshift distribution $n_i(z)$ requires calibration due to the limit of the photo-z techniques at the current stage. The fiducial calibration method for KiDS450 and KV450 is the DIR (direct calibration, with spectroscopic redshift) method. After the calibration, the redshift distribution $n_i(z)$ is expected to shift to a more accurate position. But due to the algorithm of SC, which requires not only the photo-z $z_B$, but also the connection between photo-z and the true redshift $p(z|z^P)$, which can not be calibrated with the DIR method, for each specific galaxy. Thus if we use the uncalibrated $p(z|z^P)$ and the calibrated $n_i(z)$ together, it will lead to some bias that is not easy to specify.
	
	Therefore, in the calculation of Eq.\,\eqref{Q} and \eqref{eta}, instead of using the PDFs given by the BPZ code for every single galaxy, we choose to use the Gaussian model Eq.\,\eqref{eq: Gaussian PDF} that we previously used, so that our SC method is self-consistent, avoiding the usage of the calibrated $n_i(z)$ and the uncalibrated $p(z|z^P)$ simultaneously. The bias introduced by the assumption of Gaussian photo-z PDF will be discussed in the next paragraph as well as multiple places later in the paper. The photo-z bias parameter $\Delta_z$ is set to be 0, while for the photo-z scatter $\sigma_z$, instead of using the commonly assumed value 0.05, we use $\sigma_z=0.082$ for KiDS450 and $\sigma_z=0.061$ for KV450, borrowing the results from \cite{Wright2018}. For future photo-z techniques when the estimation for the PDFs and the best-fit ($z_B$ from BPZ for example) are more accurate, we can directly use the PDFs instead of applying such a photo-z model.
	
	\begin{figure*}
		\includegraphics[width=2.0\columnwidth]{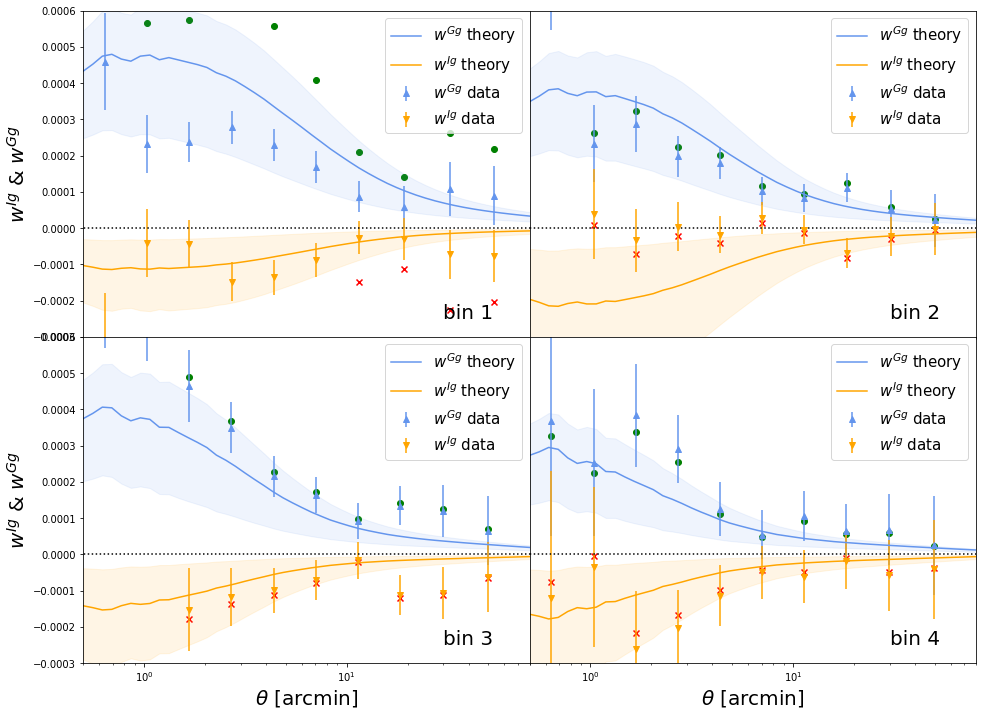}
		\caption{This figure shows, for KiDS450 data, the measurements of the pure lensing signal $w^{Gg}$ and the pure IA signal $w^{Ig}$ for the four redshift bins, presented in the four panels with the bin number in the labels. The blue up-triangles are the measured $w^{Gg}$ signals, and the yellow down-triangles are the measured $w^{Ig}$ signals, using Eq.\,\eqref{Gg correlation} and \eqref{Ig correlation} with measurements shown in Fig.\,\ref{fig: Q_i} and \ref{fig: observables}. The blue curve and the yellow curve are the theoretical predictions with KiDS450 best-fit cosmology shown in Table.\,\ref{table: fiducial cosmology}, with the DIR redshift distribution. The blue shaded area shows the upper- and lower-limits for the lensing signal under the constraint $\sigma_8=0.826^{+0.115}_{-0.199}$ from KiDS450 cosmic shear \citep{Hildebrandt2016}. The orange shaded area shows the upper- and lower-limits for the IA signal under the constraint of both $\sigma_8$ and $A_{\rm IA}=1.14^{+0.65}_{-0.55}$. The disagreements between the measurements and the theoretical predictions are mainly due to photo-z quality (which can be shown with the improvement with KV450 data later). Also, we point out that by comparing the two low-z bins and two high-z bins, as the photo-z outlier rate decreases, the agreement tends to improve. The systematical bias lead by the photo-z PDF outlier (as shown in Fig.\,\ref{fig: 2Gaussian} and \ref{fig: Q_i}) is also quantified with the green dots and red crosses, with the corresponding outlier rate [0.4, 0.15, 0.2, 0.15] for each redshift bin.}
		\label{fig: Gg and Ig}
	\end{figure*}
	
	With the above information of photo-z distribution and PDFs from data, we are able to calculate the quantity $Q_i$, which is shown in Fig.\,\ref{fig: Q_i} and Fig. \,\ref{fig:Q_i_v2}, for the lensing-IA separation later. The index $i$ is for photo-z binning, same as in KiDS450 \cite{Hildebrandt2016}, namely the $0.1<z^P<0.3$, $0.3<z^P<0.5$, $0.5<z^P<0.7$ and $0.7<z^P<0.9$ for the four redshift bins. The shaded areas are the angular scale cut, so that we only use $50<\ell<3000$, which is our main interest for cosmic shear in the current stage. This range is a little larger than the range of stage III surveys as we want to have a preview of the behavior of IA. There is some angular dependency on $\ell$ for $Q_i$, while this dependency is more and more insignificant as we go to higher redshift, or higher bin index $i$. We also showed in \cite{Yao2018} that the minimum and maximum values for $Q_i$ are very close in the same redshift bin for future surveys like LSST. Therefore in this work, we use the averaged value of $Q_i$ for the purpose of lensing-IA separation. The average value is $\bar{Q_i} \approx$ [0.588, 0.557, 0.615, 0.650] for KiDS450.
	
	According to \cite{Wright2018}, there is still a $10\%-20\%$ redshift outlier problem that can not be addressed by the above Gaussian PDF model. This outlier will reach to redshift range far outside the chosen bin, and is expected to lead to some bias in the $Q_i$ estimation by affecting $\eta_i(z)$ in Eq.\,\eqref{eta}, as well as the measurement of power spectra (2-points correlation function in this work, see the next section) in Eq.\,\eqref{Gg} and \eqref{GgS}. 
	To quantify the potential bias from the photo-z outlier, a bi-Gaussian photo-z PDF model Eq.\,\eqref{2G model} is applied, beyond the default single-Gaussian model Eq.\,\eqref{eq: Gaussian PDF}. The bi-Gaussian distributions are shown in Fig.\,\ref{fig: 2Gaussian}, and the resulting $Q_i$ values are shown in the right panel of Fig.\,\ref{fig: Q_i}. We see that the systematic bias in $Q_i$ due to the photo-z outlier is not significant, when the outlier rate is $<\sim20\%$, as in Bin 2 to 4. Later in this paper, we will also propagate this potential bias from the $Q_i$ measurement to the two-point statistics.
	
	To limit the impact of photo-z quality, we also applied a wider selection of the photo-z bins in order to alleviate the problem. For future surveys, as the error in photo-z decreases, the redshift scatter $\sigma_z$ will be smaller and the outlier rate will also decrease. So we broaden the redshift bins in KiDS450 data to simulate this improvement in the future. The redshift bins after the broadening are $0.1<z^P<0.5$ and $0.5<z^P<0.9$, with the associated values $\bar{Q_i}\approx$ [0.655, 0.674]. We then apply the same binning to KV450 data to see the improvements with better photo-z \cite{Wright2018}, with the resulting values $\bar{Q_i}\approx$ [0.638, 0.574]. The improvements will be shown together with the measurements of the correlation functions in Section \ref{Section results KiDS}.

	

	
	
	\begin{figure}
		\includegraphics[width=1.0\columnwidth]{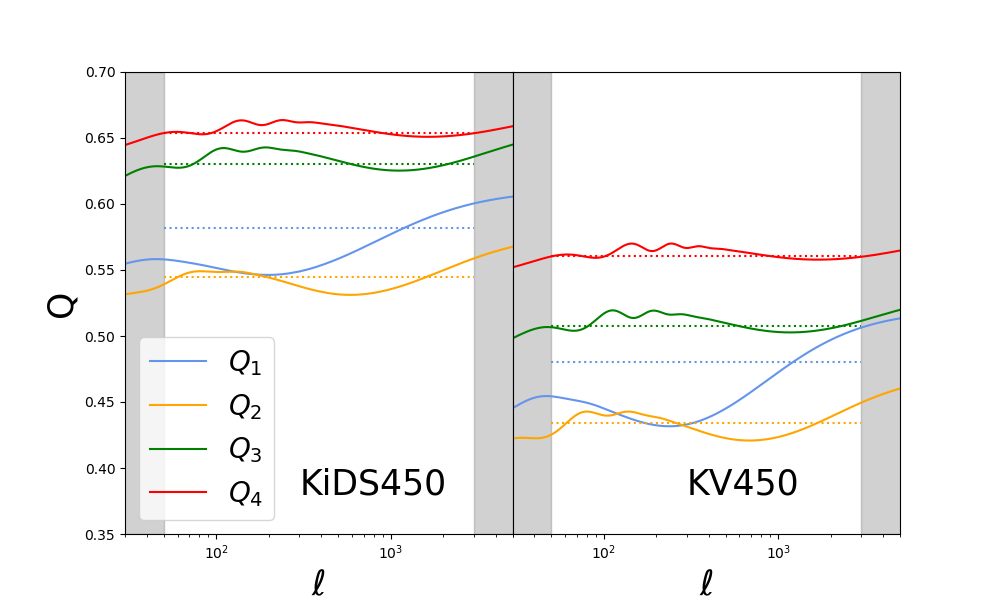}
		\caption{We show the $Q_i(\ell)$ calculated for the 4 redshift bins of KV450, as an improvement from KiDS450. The gray shaded regions are the cut in $\ell$, so that we only account for $50<\ell<3000$ in this work. The curves has weak angular dependency on $\ell$, and is nearly constant as it goes to higher redshift. The dotted lines are the mean $Q_i$ values in the 4 redshift bins, with $\bar{Q_i} \approx$ [0.58, 0.54, 0.63, 0.65] (for the default 1-Gaussian model) for KiDS450 redshift and $\bar{Q_i} \approx$ [0.48, 0.43,  0.51, 0.56] for the KV450 redshift. The clear decrease in $Q_i$ from KiDS450 to KV450 demonstrated the statistical improvement in photo-z quality.}
		\label{fig: Q_i KV 4 bins}
	\end{figure}
	
	\begin{figure}
		\hspace{-0.3cm}
		\includegraphics[width=1.1\columnwidth]{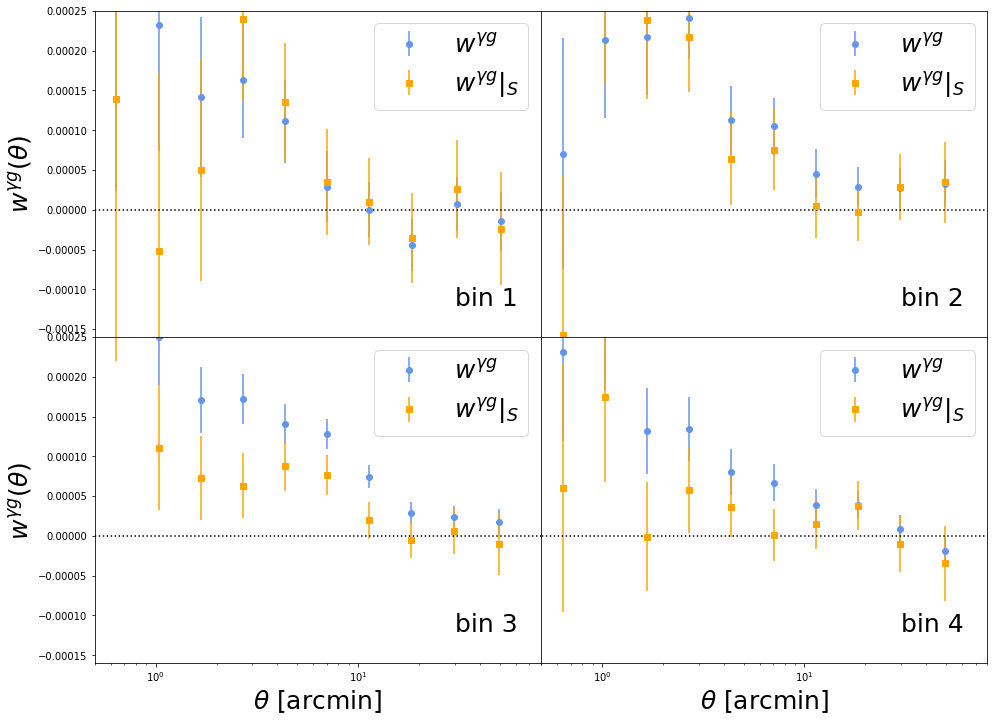}
		\caption{The measurements of the observables of SC, in the four redshift bins with KV450 data. The shape-galaxy correlations $w^{\gamma g}$ are shown in blue dots, and the shape-galaxy correlations with the selection $w^{\gamma g}|_S$ are shown in yellow squares. The redshift bin numbers from 1 to 4 are in the labels. For the two high-z bins there is a clear separation between the two correlation functions, representing the drop of the lensing signal related with Eq.\,\eqref{Q}, due to the selection of Eq.\,\eqref{selection}. But in the two low-z bins the lensing-drop due to the selection is very insignificant, as in KV450 the two low-z bins have a much smaller number of galaxies. The statistical differences for $w^{\gamma g}$ and $w^{\gamma g}|_S$ are [$2.6\sigma$, $2.9\sigma$, $6.4\sigma$, $5.1\sigma$] for each redshift bin. Since the separation for the two low-z bins are quite low, they are not suitable for applying the lensing-IA separation.}
		\label{fig: observables KV}
	\end{figure}
	
	\subsection{Correlation functions} \label{Section results 2pt} 
	
	We perform the measurements of the correlation functions as described in Eq.\,\eqref{gamma-g correlation}, for both the shape-galaxy correlation $w^{\gamma g}$ and the one with the selection $w^{\gamma g}|_S$, where the selection function is given by Eq.\,\eqref{selection}. We included a cartoon in Appendix.\,\ref{Apdx: cartoon} to explain the physical meaning behind this selection, in support of the equations.
	
\begin{figure*}
		\includegraphics[width=2.0\columnwidth]{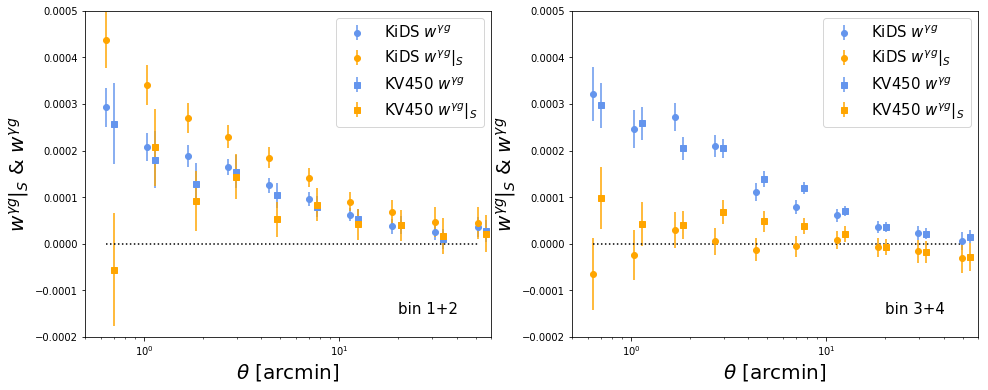}
		\caption{The measurements of $w^{\gamma g}$ and $w^{\gamma g}|_S$ for both KiDS450 and KV450, with wider photo-z bins. The bin numbers in the labels ``1+2'' and ``3+4'' are the combined redshift bins, corresponding to $0.1<z^P<0.5$ and $0.5<z^P<0.9$, respectively. In the right panel for the high-z bin, the blue $w^{\gamma g}$ dots moved to the blue squares when switching to KV450 from KiDS450, meanwhile the orange $w^{\gamma g}|_S$ dots moved to the orange squares. This is the result of using better photo-z. In the left panel for the low-z bin, similar things happened, however, the photo-z error dominates the measurement.}
		\label{fig: KiDS and KV}
	\end{figure*}

	We use the TreeCorr code \cite{Jarvis2004} to calculate the correlations in Eq.\,\eqref{gamma-g correlation}, in order to get \eqref{Ig correlation}. To account for the shape noise and sample variances correctly, we use the jackknife resampling method for this purpose. We applied 453 jackknife regions, using each KiDS450 tile as a jackknife region \citep{Hildebrandt2016}. The estimated size of a jackknife region is at $\sim1$ deg level. To avoid the edge effect of the jackknife regions \citep{Mandelbaum2006}, we calculate the correlation function $0.5<\theta<60$ arcmin, dividing into 10 logarithm bins. Each result in the Figs.\,\ref{fig: observables}, \ref{fig: Gg and Ig}, \ref{fig: KiDS and KV} and \ref{fig: KV fits} in this work has a specific covariance matrix, produced using the jackknife resampling, to calculate the errorbar on the quantity.
	
	The results of $w^{\gamma g}$ and $w^{\gamma g}|_S$ measurements are shown in Fig.\,\ref{fig: observables}. The blue dots contain $w^{\gamma g}=w^{Gg}+w^{Ig}$, while the orange squares contain $w^{\gamma g}|_S\approx\bar{Q_i}w^{Gg}+w^{Ig}$. The approximation is due to assuming $Q_i$ is angular scale independent, so that $\bar{Q_i}$ is used, according to Fig.\,\ref{fig: Q_i}. The drop in the signal in $w^{\gamma g}|_S$ is caused by the selection of Eq.\,\eqref{selection} so that some lensing signal is lost.
	
	In Fig.\,\ref{fig: observables} there is a clear separation between $w^{\gamma g}$ and $w^{\gamma g}|_S$, which works as a first proof that the physical differences between the lensing signal and the IA signal can be distinguished with the selection Eq.\,\eqref{selection} in the SC method with the current stage KiDS real data. We quantify the significance of the lensing-drop from $w^{\gamma g}$ to $w^{\gamma g}|_S$ for each redshift bin is [$8.7\sigma$, $8.3\sigma$, $9.0\sigma$, $4.5\sigma$], considering the full covariance matrix as discussed in \cite{Yao2020}. The measurements of the pure lensing signal $w^{Gg}$ and the pure IA signal $w^{Ig}$ can be therefore achieved linearly according to Eq.\,\eqref{Gg correlation} and \eqref{Ig correlation}, with the measurement of $Q_i$ shown in Fig.\,\ref{fig: Q_i}.
	
	For sanity check we also produced the cross-shear measurement (by rotating the shape of galaxies with 45 deg and measuring Eq.\,\eqref{gamma-g correlation} to get the cross-shear $\gamma^x$ instead of the tangential shear $\gamma^+$).
	Our cross-shear measurements are consistent with 0 for correlations both without the selection and with the selection. Besides the first pipeline (developed by JY) which is used to produce the full results of this work, we also have a second pipeline (developed by EP) to double-check the KiDS450 part. Details about the sanity check and different pipelines are shown in Appendix \ref{Apdx:Twopipelines}.
	
	\subsection{Lensing-IA separation for KiDS450} \label{Section results KiDS}
	
	For KiDS450 data, the measurements of the separated lensing signal and IA signal are shown in Fig.\,\ref{fig: Gg and Ig}. The triangles are the measured values and the curves are the theoretical predictions. 
	
	The measurements from data are generally consistent with the theoretical curves, for both lensing and IA parts, showing that SC can already work with the data quality of the stage III surveys (KiDS, DES, HSC), and confirming that the non-linear tidal alignment model is good enough for the current surveys. Meanwhile, there are several important details that need to be discussed. They include: the different approaches in getting Fig.\,\ref{fig: Gg and Ig}, the quality of photo-z, and the galaxy bias $b_g$ for the KiDS galaxy samples.
	
	The theoretical values in Fig.\,\ref{fig: Gg and Ig} are calculated as described in Section \ref{Section theory}, while the cosmological parameters are obtained through the cosmic shear observables \citep{Hildebrandt2016}, together with other nuisance parameters for different systematics (such as IA) with specific modeling. In contrast, our SC method uses the shape-galaxy correlation but with different selections to directly measure the lensing-galaxy correlation and IA-galaxy correlation, without any strong assumptions on the IA model. The cosmic shear measurements also use information in both auto-correlation and cross-correlation for different redshift bins. The SC method, for the purpose of signal separation on the other hand, only uses shape-galaxy correlation in the same redshift bin, as described in Eq.\,\eqref{scaling-2}, \eqref{Ig correlation} and \eqref{Gg correlation}. Despite these differences, the results in Fig.\,\ref{fig: Gg and Ig} still shows good agreements between the measurements and theoretical predictions. It is not only a sanity check for us, but also a confirmation for KiDS450 cosmic shear results since our approaches are totally different.
	
	We further discuss the systematics of the SC method, starting with the impact of photo-z quality. As discussed in previous work \citep{Yao2017}, the SC method is more sensitive to photo-z quality, compared with the conventional marginalization method with IA models. This is because SC uses extra information from photo-z when applying the selection Eq.\,\eqref{selection}. Because of this sensitivity, the error in photo-z (for example, the catastrophic photo-z error which will result in the redshift outlier problem in the PDF $p(z|z^P)$ and the redshift distribution $n_i(z)$ \cite{SC2008}) could potentially be propagated, and should already be included in the measurements $Q_i$ (Fig.\,\ref{fig: 2Gaussian}) and therefore in the two-point statistics (Fig.\,\ref{fig: Gg and Ig}). The bias introduced by the photo-z outlier is also quantified in Fig.\,\ref{fig: Gg and Ig}, by using our bi-Gaussian model in Fig.\,\ref{fig: 2Gaussian} \& \ref{fig: Q_i}. We found that for photo-z outlier rate $<0.2$ (except for the 1st bin), the introduced bias is significantly smaller than the current statistical error.
	Right now the measurements generally agree with the theoretical prediction, but for future surveys with much larger numbers of galaxies, the errorbar will shrink and such biases could appear. The bias originating from photo-z error will be addressed later in this paper when we switch to KV450 for better photo-z \citep{Wright2018}.
	
	More specifically, the two observables $w^{\gamma g}$ and $w^{\gamma g}|_S$ have different sensitivity to photo-z due to the selection Eq.\,\eqref{selection}. Thus their measurements in Fig.\,\ref{fig: observables} will be biased differently. So for better photo-z, $w^{\gamma g}$ and $w^{\gamma g}|_S$ will shift to different positions, leading to more accurate values in Fig.\,\eqref{fig: Gg and Ig}. This will also be demonstrated later with KV450 data. 
	
	Moreover, it is obvious that in Fig.\,\ref{fig: Gg and Ig}, the agreement from the two high-z bins is better than that of the two low-z bins. We think this is also caused by the photo-z outliers. As discussed in \cite{Hildebrandt2016}, the redshift outlier fraction is higher at lower redshift, leading to some bias in Fig.\,\ref{fig: observables}, which is then propagated to Fig.\,\ref{fig: Gg and Ig}. The above situations will be improved in the future with better photo-z, as shown in discussions with KV450 data.

	\begin{figure}
		\includegraphics[width=1.0\columnwidth]{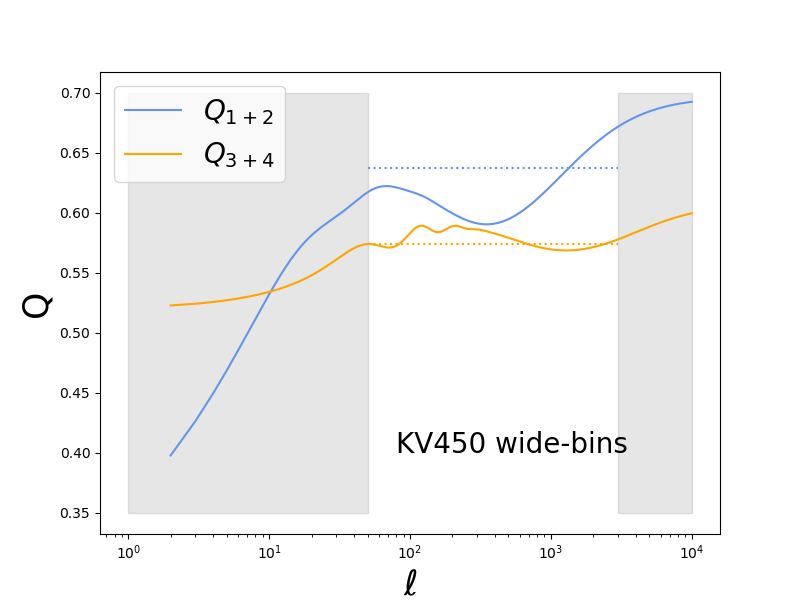}
		\caption{In this figure we show the $Q_i(\ell)$ quantity in solid curves for KV450 data. With the averaged $\hat{Q}_i$ shown in dotted curves, we can achieve the separation of lensing and IA signal from the measured observables in Fig.\,\ref{fig: KiDS and KV}. All the setups are similar to Fig.\,\ref{fig: Q_i}.}
		\label{fig: Q_i KV 2bins}
	\end{figure}
	
	Another potential bias for SC is the galaxy bias $b_g$ in Eq.\,\eqref{Gg} and \eqref{C^Ig}. The galaxy bias is the bias when using galaxy number density contrast $\delta_g$ to trace the true matter density contrast $\delta_m$, namely $\delta_g=b_g \delta_m$. In this work, we use $b_g=1$ for the theoretical predictions, for the KiDS450 galaxies. This is a reasonable assumption as the KiDS shear galaxies are not a specific kind of galaxy. But this is a bias on the modeling side. On the observational side, the galaxy bias enters both $w^{Gg}$ and $w^{Ig}$ in the same way. So one can either tune $b_g$ in the model to fit $w^{Gg}$ and $w^{Ig}$ simultaneously, or only use the IA-lensing ratio $w^{Ig}/w^{Gg}$ in which the $b_g$ cancels. Further discussion of the galaxy bias is beyond the scope of this paper.
	
	\subsection{Lensing-IA separation for KV450} \label{Section results KV}
	
	\begin{figure*}
		\includegraphics[width=2.0\columnwidth]{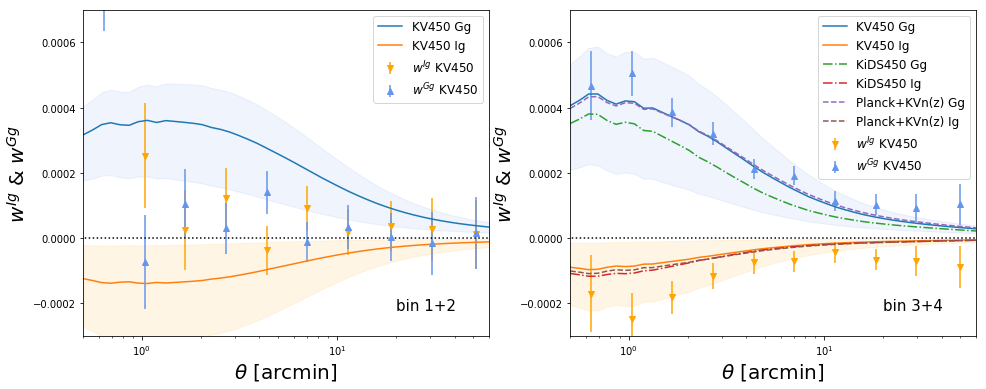}
		\caption{This figure shows the measurements of $w^{Gg}$ and $w^{Ig}$ using KV450 data and the comparison with the theoretical predictions. The blue up-triangles are the $w^{Gg}$ measurements and the orange down-triangles are the $w^{Ig}$ measurements. The theoretical values are plotted in curves with the same color, using the KV450 best-fit cosmology shown in Table.\,\ref{table: fiducial cosmology}. The blue shaded area shows the upper- and lower-limits for the lensing signal under the constraint $\sigma_8=0.836^{+0.132}_{-0.218}$ from KV450 cosmic shear \citep{Hildebrandt2018}. The orange shaded area shows the upper- and lower-limits for the IA signal under the constraint of both $\sigma_8$ and $A_{\rm IA}=0.981^{+0.694}_{-0.678}$. In the right panel, for the high-z bin, we also show some extra curves for comparison. The theoretical predictions using KiDS450 cosmology and DIR redshift distribution are also plotted in the dash-dotted curves. The ones with Planck2018 cosmology \citep{Planck2018I} plus KV450 DIR redshift distribution are plotted in the dashed curves. In the left panel the measurements from data are biased due to insignificant $w^{\gamma g}-w^{\gamma g}|_S$ separation as shown in Fig.\,\ref{fig: KiDS and KV}.}
		\label{fig: KV fits}
	\end{figure*}
	
	We now switch to KV450 data for better photo-z \citep{Wright2018} and perform comparisons to KiDS450. With ``9 KV450 bands'' photo-z rather than the ``4 KiDS450 bands'', the best-fit BPZ $z_B$ is expected to be more accurate. However, due to better photo-z quality, a lot of galaxies which were originally in the low-z bins of KiDS450 now belong to the high-z bins of KV450. This lead to a significant decrease in the available galaxy number density for the low-z bins of KV450. We calculated the ratio between the number of KV450 galaxies and the number of KiDS450 galaxies, namely $N_g^{\rm KV450}/N_g^{\rm KiDS450}$, which gives [0.32, 0.66, 1.16, 1.04] for the four redshift bins being used. The measurements for the two observables $w^{\gamma g}$ and $w^{\gamma g}|_S$ for KV450 are shown in Fig.\,\ref{fig: observables KV}. Due to the low separation-significance (resulting from the significantly lower galaxies numbers) for the two low-z bins, we decide this binning is not suitable for KV450.
	
	We further use wider photo-z bins, to anticipate future photo-z quality when the redshift PDFs are more concentrated compared with the bin-width. The new redshift bins are $0.1<z^P<0.5$ and $0.5<z^P<0.9$, i.e. we combine the original 2 low-z bins, and the 2 high-z bins. This will also increase the number of galaxies in each redshift bin, offering stronger constraining power, which is also similar to future surveys.
	
	In Fig.\,\ref{fig: KiDS and KV} we present the measurements of $w^{\gamma g}$ and $w^{\gamma g}|_S$, which is similar to Fig.\,\ref{fig: observables} \& \ref{fig: observables KV}, but with the wider redshift bins. The measurements for both KiDS450 and KV450 are plotted, to show the improvement with KV450 photo-z. In the right panel, for the high-z bin, both $w^{\gamma g}$ and $w^{\gamma g}|_S$ are shifted when switching to KV450 from KiDS450. The shift is generally larger for $w^{\gamma g}|_S$ (orange dots shift to orange squares), compared with the shift of $w^{\gamma g}$ (blue dots shift to blue squares). This is due to the selection Eq.\,\eqref{selection} using photo-z so that $w^{\gamma g}|_S$ will be affected more when changing the quality of photo-z.
	
	On the other hand, in the left panel, for the low-z bin, we can observe a similar shift. However, the results are strongly biased, as for KiDS450, the orange $w^{\gamma g}|_S$ is higher than the blue $w^{\gamma g}$, which is different from our expectation, which is, after applying the selection of Eq.\,\eqref{selection}, there should be a lensing-drop so that $w^{\gamma g}|_S<w^{\gamma g}$. Even after using KV450 data the blue squares and the orange squares can not be separated clearly. This, again we believe, is related to the redshift outlier problem, in the way that a large fraction of BPZ best-fit $z_B$ is misestimated. When we apply the combined wide-bins, some of the misestimated $z_B$ of bin 1 are actually in the range of bin 2 (and vice versa), causing some lensing signal to be treated as IA signal, while some IA signal is treated as lensing signal. Therefore the selection didn't cause a ``lensing-drop'', however with an increase in $w^{\gamma g}|_S$. This will lead to a biased estimation for $w^{Gg}$ and $w^{Ig}$, which is shown later in this section with Fig.\,\ref{fig: KV fits}. The impact from photo-z outliers leading to a biased estimation of the lensing and IA signal is also discussed in KiDS450 and CFHTLenS in \cite{Hildebrandt2016}.
	
	To properly distinguish the lensing signal and the IA signal, we recalculated the $Q_i$ values with the new binning method, similar to the steps to produce Fig.\,\ref{fig: Q_i}. We use $\bar{Q_i}\approx$ [0.655, 0.674] for KiDS450 and  $\bar{Q_i}\approx$ [0.638, 0.574] for KV450, as discussed in Section \ref{Section Q}. The $Q_i$ values for KV450 are shown in Fig.\,\ref{fig: Q_i KV 2bins} and will be used for the lensing-IA signal separation.
	
	To produce the theoretical prediction for $w^{Gg}$ and $w^{Ig}$ for the wide bins, the redshift distribution is also required. We use the galaxy number counts in each redshift bin $N_g=$ [3879823, 2990099, 2970570, 2687130] \citep{Hildebrandt2016} as weights to get the new weight-stacked DIR redshift distribution for KiDS450, and similarly $N_g=$ [1253582, 1985201, 3450970, 2792105] is used for KV450 \citep{Hildebrandt2018}. The highest-z bin $0.9<z^P<1.2$ from KV450 is dropped as it has no counterpart in KiDS450 to compare with.
	
	With the $Q_i$ shown in Fig.\,\ref{fig: Q_i KV 2bins} and the correlation functions shown in Fig.\,\ref{fig: KiDS and KV}, the pure $w^{Gg}$ and $w^{Ig}$ can be separated, presented in Fig.\,\ref{fig: KV fits}. In the left panel of Fig.\,\ref{fig: KV fits} the detection is strongly biased. This is caused by the impact of significant photo-z outliers, such that the low-z bin in Fig.\,\ref{fig: KiDS and KV} is strongly biased, which is propagated to the lensing-IA separation in Fig.\,\ref{fig: KV fits}, as discussed above.
	
	In the right panel of Fig.\,\ref{fig: KV fits}, we achieved a very good agreement between the measurements and the theoretical values. The shaded regions are when allowing $\sigma_8$ and $A_{\rm IA}$ to vary in the $1-\sigma$ range given by \cite{Hildebrandt2018}, to see how good the results of high-z bin using SC can fit into the full cosmic shear of KV450. The lensing signal has a very good agreement with the prediction, which stands as another sanity check. The IA signal, however, suggests a larger $A_{\rm IA}$ compared with the KV450 best-fit of cosmic shear. 
	
	The correlation functions with KiDS450 cosmology and redshift distribution, as well as the ones with Planck cosmology plus KV450 redshift distribution, are also presented for comparison. The agreement between our measured $w^{Gg}$ using SC and the theoretical values of KV450, is generally similar to the agreement between KiDS450 and KV450. We can also see that even though the best-fit cosmology of KiDS450 and KV450 are closer, the theoretical $w^{Gg}$ in Fig.\,\ref{fig: KV fits} agrees better for KV450 and Planck, as these two cases both use the DIR redshift of KV450. This again emphasizes the importance of the impact of photo-z.
	
	By comparing the high-z bin ($0.5<z^P<0.9$) of Fig.\,\ref{fig: KV fits} and the counterparts in Fig.\,\ref{fig: Gg and Ig}, we can see the improvements of SC in the future as data quality improves. First of all, the error bar will be significantly reduced due to more galaxies being observed. Secondly, with better photo-z, the bias from photo-z in the measurements of $w^{Gg}$ and $w^{Ig}$ will be reduced, giving more accurate results. This improvement in the accuracy is expected to be better compared to the marginalization method as SC is more sensitive to photo-z, as shown in this work and discussed previously \citep{Yao2017}.
	
	\section{Summary and conclusions} \label{Section discussion}
	
	In this paper, we presented the first application of the self-calibration (SC2008, i.e. \cite{SC2008,Yao2017,Yao2018}) method to distinguish the galaxy lensing-galaxy density (Gg) signal and the IA-galaxy density (Ig) signal with the shear catalog from real data with photo-z. We developed the applicable steps on how the SC method can be used. We apply the steps to KiDS450 data as a first approach, and later apply them to KV450 data to discuss the impact of photo-z quality. We take systematics such as the shape calibration, selection bias, survey shape, photo-z error, and galaxy bias into consideration. We use jackknife resampling to produce the covariance matrix, in order to calculate the error estimation. We perform several sanity checks such as cross-shear measurements, different pipelines (Appendix \ref{Apdx:Twopipelines}),  and comparing the lensing signals from SC with those from theoretical predictions extrapolated using the best-fit cosmology from cosmic shear, with the same data sets. We showed a good agreement using KiDS450 data, for both lensing and IA measurements. We further extend our analysis to KV450, focusing on $0.5<z^P<0.9$. For the selected KV450 data, the lensing measurements are consistent, while the results in IA suggest a stronger amplitude compared with \cite{Hildebrandt2018}. We think this is mainly due to: (1) we are only using the high-z samples, which is expected to have a different IA amplitude; (2) we are not including the effect of galaxy bias, which is degenerate with the IA amplitude in the $w^{Ig}$ signal at some level; (3) the photo-z error could still lead to some bias in our analysis.
	
	To further extend the SC method, we discussed several systematic effects that can potentially bias the lensing-IA separation, including:\\
	$\bullet$ The approximation of using the averaged $\bar{Q_i}$ instead of a scale-dependent $Q_i(\ell)$. This is not a severe problem, because $Q_i(\ell)$ will behave more like a constant with higher redshift. Also, the scale-dependent $Q_i(\ell)$ will shift power in between large- and small-scales, leading to a tiny ``slope-change'', while the overall IA amplitude should not be biased significantly. Even so, it is still worth exploring for the relatively shallow surveys. For related discussions, see \cite{Yao2020}. \\
	$\bullet$ The photo-z error (mainly the redshift outlier). This includes both the error in the photo-z PDFs $p(z|z^P)$ and the error in the best-fit photo-z ($z_B$ for BPZ code for example). As discussed in the paper, the error in $p(z|z^P)$ can potentially bias the estimation of $Q_i$ (Fig.\,\ref{fig: Q_i}) and the resulting correlations (Fig.\,\ref{fig: Gg and Ig}); while the error in best-fit $z^P$ can bias the measurements of the two correlation functions $w^{\gamma g}$ and $w^{\gamma g}|_S$, as shown in Fig.\,\ref{fig: KiDS and KV} for the similar galaxy samples but with different redshift measurements. The error from the biased $z_B$ is included in our jackknife resampling. The photo-z error can lead to significant bias in SC, and with the improvement of photo-z techniques, this bias is expected to drop quickly due to the sensitivity discussed in both this work and \cite{Yao2017}. In the future, it is also possible to combine photo-z surveys and spectroscopic redshift surveys to obtain a more reliable PDF $p(z|z^P)$, then use the Monte Carlo sample from the PDF to obtain a more accurate $Q_i$ value.\\
	$\bullet$ The galaxy bias $b_g$. In this work, we showed the assumption of $b_g=1$ is good enough for the KiDS450 and KV450 shear galaxy samples. To avoid this problem we can either choose to use the IA-lensing ratio instead of the signals themselves, since $b_g$ enters these two signals in the same way, or we can use what is suggested in \cite{SC2008} to directly infer it. In the future, it is worth to explore both of these two approaches. \\
	$\bullet$ We choose to ignore the impact of baryonic effects such as massive neutrinos and feedbacks, as they affect the modeling part, which is beyond the scope of this paper. But including them in future analysis will also be helpful. \\
	$\bullet$  Nonetheless, we point out here there is some (weak) potential bias for this selection method Eq.\,\eqref{selection} using photo-z of SC. Since photo-z is obtained from different filter bands, for example $u g r i$ four bands for KiDS450, the selection in photo-z could potentially lead to some selection effect in the color of the galaxies, i.e. selection effect in galaxy types. Since galaxy types are expected to be related to the IA signal, the IA signals could also be different for different selection $S=0$ or $S=1$. This selection effect is expected to be weak, but we still want to point it out for future studies.
	
	Based on the good agreement we achieved for KiDS450 and KV450 and the sanity checks we performed, we conclude that the above systematics are not severe for the current data quality. We then discuss the cosmological information that can be put into use in the future, based on our results. \\
	$\bullet$ The subtracted IA signal $w^{Ig}$ (as shown in Fig,\,\ref{fig: Gg and Ig} and \ref{fig: KV fits}) contains information of the IA, without any strong assumption on the physics behind. This information can be propagated to either cosmic shear or CMB-lensing to mitigate the IA contamination, as previously discussed \cite{SC2008,Zhang2010,Yao2017,Yao2018,Troxel2014,Meng2018}, which was our motivation of this work at the beginning. \\
	$\bullet$ The subtracted IA signal can be used to directly constrain the IA model, rather than the marginalization method that fits the cosmological parameters together with the nuisance IA parameters. We perform in Appendix \ref{Apdx: MCMC IA} an example of using the IA signal subtracted from KV450, Fig.\,\ref{fig: KV fits}, to constrain the tidal alignment model of IA, which is used in the KV450 paper. A more detailed analysis can be performed when the systematics of the SC method are fully understood. Then this information can also be used to distinguish different IA models, which are widely discussed in \cite{HSC_Hikage2019,HSC_Hamana2019,Troxel2017,Blazek2017,Krause2016}. \\
	$\bullet$ The subtracted lensing signal $w^{Gg}$ can be used to constrain cosmology. This quantity uses $w^{\gamma g}$ and $w^{\gamma g}|_S$, therefore its uncertainty is larger than both of them. However, since this measurement uses all the shear galaxies, meaning that it uses the galaxy number density $\delta_g$ as the tracer for matter, so that the overall galaxy number is much larger than the typical galaxy-galaxy lensing or galaxy auto-correlation where spectroscopic redshift is required. And this galaxy information in the shear catalog has never been used in previous cosmological studies. We performed an MCMC study for $\Lambda$CDM model in Appendix \ref{Apdx: MCMC cosmo}, without further addressing the systematics in the SC method nor the nuisance parameters for $\Lambda$CDM. It works as a first example of the extra constraining power of $w^{Gg}$ with SC. \\
	$\bullet$ The IA signal itself does contain some cosmological information, meaning that our subtracted $w^{Ig}$ can be used together with $w^{Gg}$ to constrain the IA model and cosmological model simultaneously. In this way, some of the constraining power in $w^{Ig}$ will flow from IA parameters to cosmological parameters. If the cosmological model and the IA model are accurate enough, this method is expected to be consistent with the marginalization method. \\
	$\bullet$ In this paper, we follow the process of the original SC method \citep{SC2008}, to obtain $w^{Gg}$ and $w^{Ig}$ and discussed the information that we can put into use in the future. On the other hand, we can directly use the SC observables $w^{\gamma g}$ and $w^{\gamma g}|_S$ as data, and theoretically calculate the predictions from cosmological models. This can prevent the worry about the cosmology-dependency of the $Q_i(\ell)$ we previously discussed \citep{Yao2017}. Although this dependency is expected to be very weak, as the same cosmological model enters both the numerator and denominator of Eq.\,\eqref{Q} and \eqref{eta}.
	
	We then list the ongoing and planned future work. The extension from KiDS450 to future KiDS data release will be straightforward, and we will include more detailed discussions on the systematics of SC and their corrections. Applying SC to a more wide range of redshift (HSC data) and testing different IA models (DES data) are also of great importance. We are also testing and quantifying the impact of the redshift outlier with different simulations and with machine learning photo-z.

	
	\section{acknowledgements}
	The analysis is based on data products from observations made with ESO Telescopes at the La Silla Paranal Observatory under programme IDs 177.A-3016, 177.A-3017, 177.A-3018, 179.A-2004, 298.A-5015, and on data products produced by the KiDS consortium.
	
	The authors thank Konrad Kuijken, Hendrik Hildebrandt, Rui An, Mike Jarvis, Lindsay King, Zhaozhou Li, Michael A. Troxel, and Yu Yu for useful discussions. 
	MI acknowledges that this material is based upon work supported in part by the U.S. National Science Foundation under grant AST-1517768 and the U.S. Department of Energy, Office of Science, under Award Number DE-SC0019206. The authors acknowledge the Texas Advanced Computing Center (TACC) at The University of Texas for providing HPC resources that have contributed to the research results reported within this paper. URL: http://www.tacc.utexas.edu. 
	PZ acknowledges that this work was supported by the National Science Foundation of China (11621303, 11433001, 11653003).
	The codes JY produced for this paper were written in Python. JY thanks all its developers and especially the people behind the following packages: SCIPY \citep{scipy}, NUMPY \citep{numpy}, ASTROPY \citep{astropy} and MATPLOTLIB \citep{matplotlib}.

	\bibliography{References}
	\bibliographystyle{mnras} 
	
	\appendix
	
	\section{Illustrating the SC observables} \label{Apdx: cartoon}
	
	\begin{figure}
		\includegraphics[width=1.0\columnwidth]{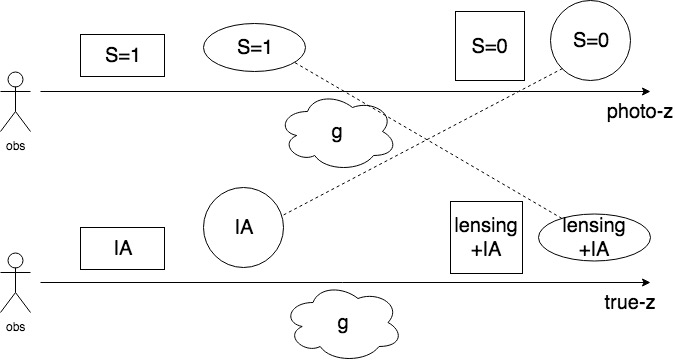}
		\caption{This is a cartoon to illustrate how the selection Eq.\,\ref{selection} of SC works. Above the redshift arrow, the four shapes (rectangle, circle, square, ellipse) are the galaxies whose shape information we use. Below the redshift arrow, the cloud represents the galaxy whose number density information (g) we use. When calculating the shape-galaxy correlation, in the lower part when considering true-z space, the correlation contains both lensing and IA for the pairs with shape galaxies farther away than the number count galaxies (square and ellipse), while it only contains IA signal for the pairs with shape galaxies closer than the number count galaxies.
			In photo-z space in the upper part, however, the ordering of the shape galaxies and the number count galaxies could be different due to the photo-z error. In the figure, following the dotted lines, the ellipse is mistaken to be closer to us, and the circle galaxy is mistaken to be farther. So without the selection, all the galaxies pairs (both $S=0$ and $S=1$) are used to get $w^{\gamma g}$, while with the selection of SC, only the $S=1$ galaxy pairs (for rectangle and ellipse) are selected to calculate $w^{\gamma g}|_S$. The only lensing signal contained in this kind of pairs comes from the ones with photo-z error (the ellipse).  Therefore, $w^{\gamma g}|_S$ will have a drop in the lensing signal, and if we know the photo-z quality we can figure out how significant this drop is, quantified by the $Q_i$ values. In this way, the true lensing signal $w^{Gg}$ and the true IA signal $w^{Ig}$ can be solved.}
		\label{fig: cartoon}
	\end{figure}
	
	Here we show a cartoon Fig.\,\ref{fig: cartoon} to give a more detailed explanation of the physics behind the $w^{\gamma g}$ and $w^{\gamma g}|_S$ measurements. For readability reasons, we include the full discussion in the caption.
	
	\section{Two-pipelines}     \label{Apdx:Twopipelines}
	
	\begin{figure}
		\includegraphics[width=1.0\columnwidth]{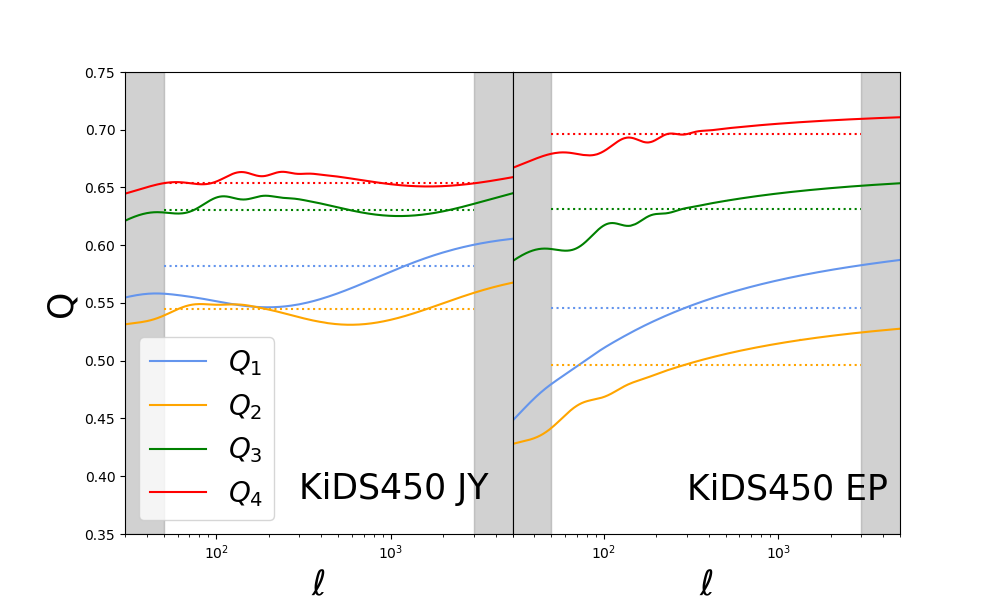}
		\caption{The comparison of $Q_i(\ell)$ calculated using two pipelines with KiDS450 data. The averaged $Q_i$ values are shown in the dotted lines. The differences between the two pipelines could be due to different target selection, different integration methods in $\eta$, Eq.\,\eqref{eta}, and the cosmology codes being used. However the biases due to this difference is expected to be insignificant, as the introduced error are similar as in Fig\,\ref{fig: Q_i} and \ref{fig: Gg and Ig}.}
		\label{fig:Q_i_v2}
	\end{figure}
	
	\begin{figure}
		\centering
		\hspace*{-0.5cm}
		\includegraphics[width=1.1\columnwidth]{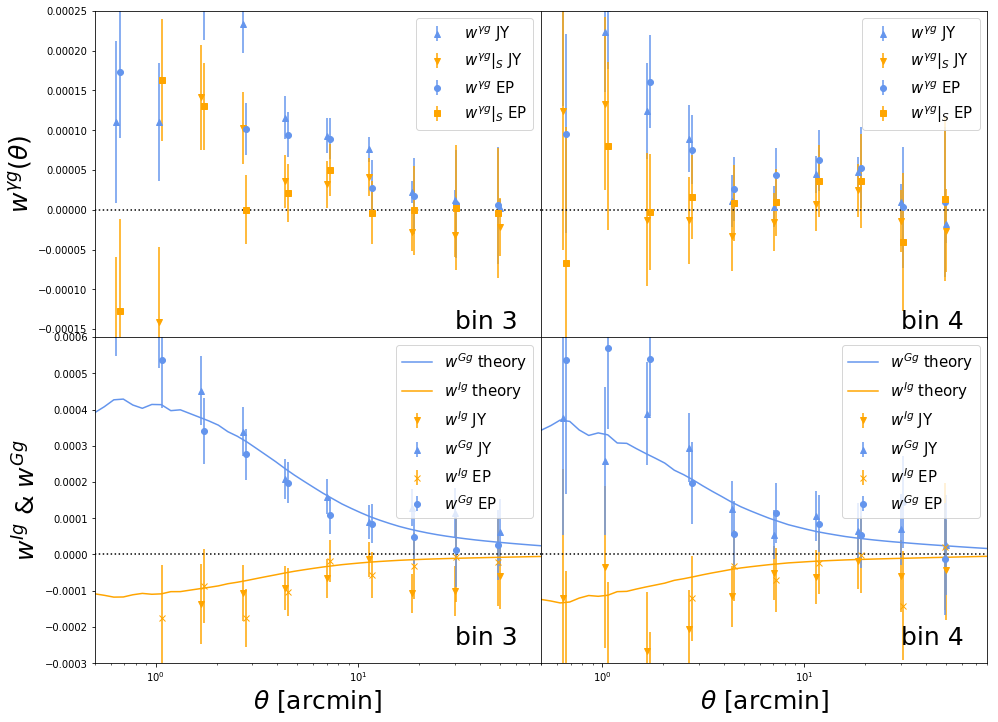}
		\caption{We show in this figure the agreement of the two pipelines in both the two observed 2-point correlation functions (dots for $w^{\gamma g}$ and squares for $w^{\gamma g}|_S$) in the upper panels, and the separated pure lensing signal (up-triangles for $w^{Gg}$) and IA signal (down-triangles for $w^{Ig}$) in the lower-panels. We show only the two high-z bins labeled as 3 and 4, as in the high-z bins the photo-z outlier problem is less significant as discussed in Fig.\,\ref{fig: Gg and Ig}. ``JY'' denotes the default pipeline, and ``EP'' denotes the 2nd pipeline. In general, the two pipelines agrees very well, with some small deviations which are caused by the detailed differences described previously and in Appendix\,\ref{Apdx:Twopipelines}, while the systematics such as the photo-z error and the shape calibration will also enter the two pipelines accordingly. The two curves are the theoretical predictions with KiDS450 best-fit cosmology shown in Table.\,\ref{table: fiducial cosmology}, with the DIR redshift distribution.}
		\label{fig:2-ppl}
	\end{figure}
	
	\begin{figure}
		\centering
		\hspace*{-0.5cm}
		\includegraphics[width=1.1\columnwidth]{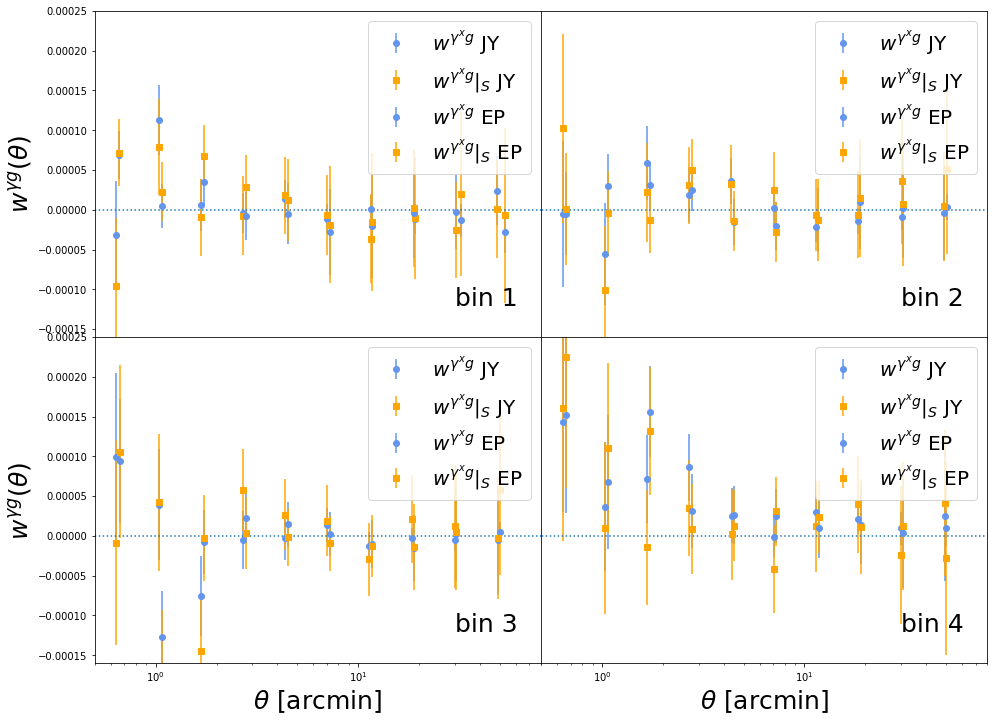}
		\caption{We show in this figure the null test of the 45-degree cross-shear correlation and the agreement of the two pipelines in both with (dots for $w^{\gamma^x g}$) and without (squares for $w^{\gamma^x g}|_S$) the SC selection of Eq.\,\eqref{selection}. ``JY'' denotes the default pipeline, and ``EP'' denotes the 2nd pipeline. In general the two pipelines agrees well, and are consistent with 0.}
		\label{fig: x-check}
	\end{figure}
	
	We applied two separate pipelines to process the KiDS450 data for rigorous reasons. The first pipeline is fully developed by JY, while the second pipeline was developed by EP. The main differences between the two pipelines include: different shape calibration and different version of TreeCorr code will lead to different correlation function measurements; different covariance matrices due to different Jackknife regions;  different numerical approaches for photo-z and $Q_i$ values.
	
	For the second pipeline, we make use of Treecorr's \citep{Jarvis2004} built-in functionality to have the correlation estimator become:
	\begin{equation}\label{estimatorpipe2}
	w^{\gamma g} = NG - RG,
	\end{equation}
	where the $NG$ refers to the number count-shape correlation for the same catalog and bin. While $RG$ is the number count-shape correlation between the random catalog and the same catalog and bin as for the shape in $NG$.
	while the shape calibration for the multiplicative bias $m$ in the second pipeline is included in the individual galaxy shape. 
	
	The second pipeline also makes use of jackknife resampling to account for the shape noise and sample variance. However, instead of using the KiDS450 tiles as regions, it makes use of the kmeans\_radec code (\url{https://github.com/esheldon/kmeans_radec}) to instead divide the sample, i.e. each tomographic bin into chunks with roughly the same number of galaxies in each chunk. 
	
	In the second pipeline, the random catalog generated using the KiDS450 footprint mask is of a fixed size (in our case $10^8$ objects) generated using healpix\_util\footnote{https://github.com/esheldon/healpix\_util}, for each bin.
	
	Different from pipeline one of this work, in the second pipeline with KiDS450 data, we handle the multiplicative bias by instead calculating the weighted average multiplicative bias for each bin, and dividing by it. 
	
	Despite the differences discussed above, the two pipelines still converge to very close results, see Fig.\,\ref{fig:Q_i_v2},\& \ref{fig:2-ppl}, demonstrating the robustness of the pipelines, as well as the stability of the IA separation method in SC \citep{SC2008} against those changes. We present here one more result of the cross-shear null test in Fig.\,\ref{fig: x-check} in support of the results in Fig.\,\ref{fig:2-ppl}. The formula being used is similar as in Fig.\,\ref{fig:2-ppl}, while the galaxy shapes are rotated by 45 degrees so that the correlated shape is the cross shear $\gamma^x$ rather than the tangential shear $\gamma^+$. The results are consistent with 0 for both pipelines and for with and without the SC selection Eq.\,\eqref{selection}.
	
	\section{Constraining the IA model} \label{Apdx: MCMC IA}
	
	\begin{figure}
		\centering
		\includegraphics[width=0.9\columnwidth]{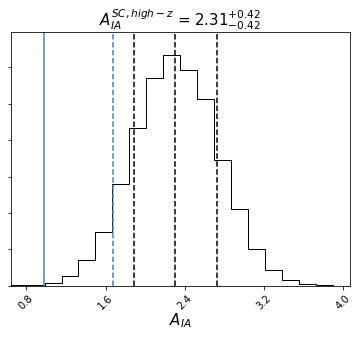}
		\caption{The MCMC constraint on the IA amplitude $A_{\rm IA}$ of the tidal alignment model Eq.\,\eqref{IA 3D}, using the IA signal subtracted in the high-z bin in Fig.\,\ref{fig: KV fits}. The y-axis is the likelihood, giving the best-fit $A_{\rm IA}=2.31^{+0.42}_{-0.42}$. The best-fit and the $\pm 1\sigma$ values are shown in the black dashed lines, while the best-fit of KV450 cosmic shear is $A_{\rm IA}=0.981^{+0.694}_{-0.678}$, shown in the blue line.}
		\label{fig: MCMC IA}
	\end{figure}
	
	We use the subtracted IA signal in Fig.\,\ref{fig: KV fits} of the high-z bin ($0.5<z^P<0.9$) to constrain the tidal alignment model as in Eq.\,\ref{IA 3D}. We use MCMC to get the constraint on the IA amplitude parameter $A_{\rm IA}=2.31^{+0.42}_{-0.42}$, which is larger than the result from KV450 cosmic shear $A_{\rm IA}=0.981$. Our likelihood is constructed as follows:
	\begin{equation}
	\mathcal{L} \propto {\rm exp}\left[-\frac{1}{2} (w^{Ig}_{\rm measured}-w^{Ig}_{\rm model}) (Cov^{Ig})^{-1} (w^{Ig}_{\rm measured}-w^{Ig}_{\rm model})\right],
	\end{equation}
	where $w^{Ig}_{\rm measured}$ is the IA measurements shown in Fig.\,\ref{fig: KV fits}; $w^{Ig}_{\rm model}$ is the theoretical prediction curve, interpolated at the same $\theta$ position as the $w^{Ig}_{\rm measured}$; $Cov^{Ig}$ is the covariance matrix that has been used to get the errorbars for the $w^{Ig}_{\rm measured}$ values, with jackknife resampling. The prior for $A_{\rm IA}$ is flat prior [-6, 6], same as in \cite{Hildebrandt2018}. In this work we use the emsemble sampler of emcee \citep{emcee}. We also point out that in our analysis we do not include the nuisance parameters such as baryonic feedback or redshift distribution uncertainty, so this is not a strictly fair comparison to KV450.
	
	Despite the potential systematics (galaxy bias, photo-z error) that was previously discussed in this paper, this small tension can suggest the deviation of the true IA model from the assumed tidal alignment model. It could also be due to the selection effect on the redshift that we used ($0.5<z^P<0.9$, which is different from KV450 cosmic shear), considering the evolution of IA when redshift is changing \citep{Yao2020}. As a result, it will be of great importance to apply future data to both the conventional marginalization method and our SC method to see how this tension will develop. The application of other IA models will also be helpful.
	
	We argue that an under-estimated $A_{\rm IA}$ will lead to an under-estimated $S_8$ using SC, which is also shown in some recent research \cite{Fluri2019} when deep-learning is used to subtract more cosmological information beyond the conventional two-point statistics. Thus the systematics in different approaches (e.g. SC or IA modeling) of IA mitigation will require careful discussion in the future.

	\section{The constraining power from lensing signal} \label{Apdx: MCMC cosmo}
	
	
	\begin{figure}
		\hspace{-0.5cm}
		\includegraphics[width=1.1\columnwidth]{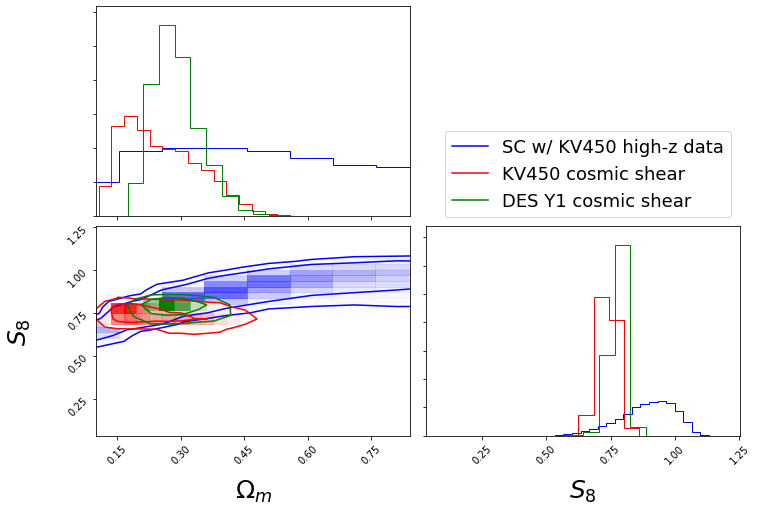}
		\caption{This figure shows the MCMC results using $w^{Gg}$ from the high-z wide-bin of KV450 in Fig.\,\ref{fig: KV fits}, to constrain $\Lambda$CDM cosmology, in particularly $S_8=\sigma_8\sqrt{\Omega_m/0.3}$ and $\Omega_m$. In the 2-D posteriors the 68\% and 95\% confidence contours are shown, with our SC results in blue, KV450 cosmic shear results in red, and DES Y1 cosmic shear results in green. We emphasis that even though our analysis used approximations for galaxy bias and $Q_i$, which could potentially lead to some small bias in the best-fit, the extra constraining power of SC-separated lensing signal and its different degeneracy direction carries valuable cosmological information.}
		\label{fig: MCMC lensing with KV and DES}
	\end{figure}
	
	This good agreement in Fig.\,\ref{fig: KV fits} suggests that we can directly use this $w^{Gg}$ lensing signal to constrain cosmology. It uses extra information in the  shear catalog that was not being applied. We can construct a likelihood for the pure lensing signal $w^{Gg}$ to constrain the cosmological parameters.
	\begin{equation}
	\mathcal{L} \propto {\rm exp}\left[-\frac{1}{2} (w^{Gg}_{\rm measured}-w^{Gg}_{\rm model}) (Cov^{Gg})^{-1} (w^{Gg}_{\rm measured}-w^{Gg}_{\rm model})\right],
	\end{equation}
	where $w^{Gg}_{\rm measured}$ is the lensing measurements shown in Fig.\,\ref{fig: KV fits}; $w^{Gg}_{\rm model}$ is the theoretical prediction curve, interpolated at the same $\theta$ position as the $w^{Gg}_{\rm measured}$; $Cov^{Gg}$ is the covariance matrix that has been used to get the errorbars for the $w^{Gg}_{\rm measured}$ values, with jackknife resampling. The priors for the cosmological parameters are the same as in \cite{Hildebrandt2018}, to prevent the impact from different priors. We note we only account for the cosmological parameters, while considering the nuisance parameters (such as feedbacks, massive neutrinos..) are beyond the scope of this paper.
	
	Our MCMC best-fit posteriors generally agree with the best-fit from KV450 cosmic shear (blue dots). The $S_8-\Omega_m$ contour is plotted together with KV450 \citep{Hildebrandt2016} and DES Y1 \citep{DES2016} results in Fig.\,\ref{fig: MCMC lensing with KV and DES}. Moreover, it is worth noticing that, despite the approximations we have made in this paper, in the $S_8-\Omega_m$ space, we have good agreements with both KV450 and DES Y1, while our degeneracy direction is different from cosmic shear. In the future when systematics are more carefully discussed, the SC-measured $w^{Gg}$ lensing signal can potentially be used to break the degeneracy.

	\label{lastpage}
\end{document}